\RequirePackage{fix-cm}
\documentclass[smallextended]{svjour3}       
\smartqed  
\usepackage{graphicx}

\usepackage{subfigure}
 \journalname{Statistical Papers}
\begin{document}

\title{On Two Simple and Effective Procedures for High Dimensional Classification of General Populations
}

\titlerunning{On Two Simple and Effective Procedures for High Dimensional Classification}        

\author{Zhaoyuan Li         \and
        Jianfeng Yao 
}


\institute{Zhaoyuan Li \at
              Department of Statistics and Actuarial Science, The University of Hong Kong, Pokfulam, Hong Kong \\
              Tel.: +852-65010250\\
              Fax: +852-28589041\\
              \email{zyli12@hku.hk}           
           \and
           Jianfeng Yao \at
              Department of Statistics and Actuarial Science, The University of Hong Kong, Pokfulam, Hong Kong\\
               \email{jeffyao@hku.hk}
}

\date{Received: date / Accepted: date}

\maketitle

\begin{abstract}
In this paper, we generalize two criteria, the determinant-based and
trace-based criteria proposed by Saranadasa (1993), to general
populations for high dimensional classification. These two criteria
compare some distances between a new observation and several different known groups. The determinant-based criterion performs well for correlated variables by integrating the covariance structure and is competitive to many other existing rules. The criterion however requires the measurement dimension be smaller than the sample size. The trace-based criterion in contrast, is an independence rule and effective in the $``$large dimension-small sample size" scenario. An appealing property of these two criteria is that their implementation is straightforward and there is no need for preliminary variable selection or use of turning parameters. Their asymptotic misclassification probabilities are derived using the theory of large dimensional random matrices. Their competitive performances are illustrated by intensive Monte Carlo experiments and a real data analysis.
\end{abstract}

\section{Introduction}
\label{intro}
In recent years, there is a great deal of attention paid to the development of high dimensional classification methods. Many independence rules are proposed to deal with the situations where the correlations between variables are weak. Tibshirani et al. (2002) proposed the nearest shrunken centroid (NSC) classifier. Fan and Fan (2008) proposed the features annealed independence rule (FAIR). Moreover, Bickel and Levina (2004) showed that the independence rule, naive Bayes (NB) performs better than the naive Fisher discriminant (NFR) where the variables are correlated. When the correlations are significant, NFR is about the same as random guess. They also showed that a classification procedure using a subset of well selected features is better than that using all the features, which typically accumulates much noise in estimating population centroids in high dimensional space.

In addition, methods integrating the covariance structure have been proposed in the literature, such as support vector machines (Vapnik 1995), shrunken centroids regularized discriminant analysis (SCRDA) (Guo et al. 2005), sparse linear discriminant analysis (Shao et al. 2011) and $D D\alpha-$procedure (Lange et al. 2014). A recent work Fan et al. (2012) proposed a new method that involves correlation information, called regularized optimal affine discriminant (ROAD). Interestingly enough, the classification error of the ROAD decreases as the correlation coefficient increases. Two variants are screening-based rules, named S-ROAD1 and S-ROAD2, which select only 10 features and 20 features, respectively. In the simulation study, under the $``$large $p$-small $n$" and equal correlation setting, the ROAD method outperforms the available classifiers mentioned above. S-ROAD2 also performs well, while S-ROAD1 fails for highly correlated variables. Notice that the ROAD and its variants have to select variables in the procedure of classification. Although variable selection has been extensively developed in last decades, their practical implementation still faces several difficult issues such as the choice of turning parameter or thresholding values. In this paper, we investigate whether there are straightforward methods that can have competitive performances without preliminary variable selection. In addition, existing methods mainly focus on $``$large $p$-small $n$" case and the localized mean vector scenario (see follows for exact definition). However, the case of $``$large $p$-large $n$" with comparable magnitude and the delocalized scenario are common issues in high dimensional classification. The classification rules proposed in the paper will help to handle these situations.

Saranadasa (1993) proposes the determinant-based (D-) and trace-based (T-) criteria. Their asymptotic misclassification probabilities are established for normal populations. In this paper, we focus on the performance of these two criteria in the delocalized scenario without the normal assumption. Specifically, consider two $p$-dimensional multivariate populations $\Pi_1$ and $\Pi_2$ with respective mean vectors $\vec{\mu}_1$, $\vec{\mu}_2$ and common covariance matrix $\mathbf{\Sigma}$. The parameters $\vec{\mu}_1$, $\vec{\mu}_2$ and $\mathbf{\Sigma}$ are unknown and thus estimated using training samples $\mathbf{X} =(\mathbf{x}_1,\mathbf{x}_2,\ldots,\mathbf{x}_{n_1})^T$ from $\Pi_1$ and $\mathbf{Y}=(\mathbf{y}_1, \mathbf{y}_2,\ldots, \mathbf{y}_{n_2})^T$ from $\Pi_2$ with respective sample size $n_1$ and $n_2$. A new observation vector, $\mathbf{z}=(z_1, z_2, \ldots, z_p)^T$ is known to belong to $\Pi_1$ or $\Pi_2$ and the aim is to find exactly its origin population. More complicated sample setting can refer to Leung (2001), which considers mixed continuous and discrete variables in each group. Cheng (2004) studies the situation where the two populations have different covariance matrices. Krzy\'{s}ko and Skorzybut (2009) considers the multivariate repeated measures data with Kronecker product covariance structures.

Let $(\bar{x})$, $(\bar{y})$ be the two training sample mean vectors where
 \begin{eqnarray*}
\bar{x}_l=\frac{1}{n_1}\sum_{i=1}^{n_1}x_{il}\quad \textrm{and}  \quad \bar{y}_l=\frac{1}{n_2}\sum_{j=1}^{n_2}y_{jl}, \quad l=1,2,\ldots,p.
\end{eqnarray*}
If the vector $\mathbf{z}$ is classified to the population $\Pi_1$, then the overall within group sum of squares and cross products matrix is
\begin{eqnarray*}
\mathbf{A}_1 = \sum_{i=1}^{n_1} (\mathbf{x}_i-\bar{\mathbf{x}})(\mathbf{x}_i-\bar{\mathbf{x}})^\prime + \sum_{j=1}^{n_2} (\mathbf{y}_j-\bar{\mathbf{y}})(\mathbf{y}_j-\bar{\mathbf{y}})^\prime + \frac{n_1}{n_1+1}(\mathbf{z}-\bar{\mathbf{x}})(\mathbf{z}-\bar{\mathbf{x}})^\prime.
\end{eqnarray*}
While, if $\mathbf{z}$ is classified to $\Pi_2$, then the sum is
\begin{eqnarray*}
\mathbf{A}_2 = \sum_{i=1}^{n_1} (\mathbf{x}_i-\bar{\mathbf{x}})(\mathbf{x}_i-\bar{\mathbf{x}})^\prime + \sum_{j=1}^{n_2} (\mathbf{y}_j-\bar{\mathbf{y}})(\mathbf{y}_j-\bar{\mathbf{y}})^\prime + \frac{n_2}{n_2+1}(\mathbf{z}-\bar{\mathbf{y}})(\mathbf{z}-\bar{\mathbf{y}})^\prime.
\end{eqnarray*}
Intuitively, one would decide $\mathbf{z}\in \Pi_1$ when $\mathbf{A}_1$ is in some sense $``$smaller" than $\mathbf{A}_2$. The D-criterion defines this smallness to be
\begin{eqnarray}
\textrm{det}(\mathbf{A}_1)<\textrm{det}(\mathbf{A}_2),\label{e1}
\end{eqnarray}
and the T-criterion defines it to be
\begin{eqnarray}
\textrm{tr}(\mathbf{A}_1)<\textrm{tr}(\mathbf{A}_2).\label{e2}
\end{eqnarray}

Two scenarios of mean difference $\vec{\delta}=\vec{\mu}_2 -\vec{\mu}_1$ are defined as follows:
\begin{enumerate}
\item \emph{Localized scenario}:  the difference $\vec{\delta}$ is concentrated on a small number of variables. We set $\vec{\mu}_1 =\mathbf{0}$ and $\vec{\mu}_2$ equals to a sparse vector: $\vec{\mu}_2=(\mathbf{1}_{n_0}^\prime, \mathbf{0}_{p-n_0}^\prime)$, where $n_0$ is the sparsity size. Notice that the location of the $n_0$ non-zero components does not influence the performance of various classifiers.

\item \emph{Delocalized scenario}: the difference $\vec{\delta}$ is dispersed in most of the variables. To ease the comparison with the localized scenario, we choose the parameters such that the averaged Mahalanobis distances are the same under these two scenarios. This is motivated by the fact that following Fisher (1936), the difficulty of classification mainly depends on the Mahalanobis distance $\Delta^2 =\vec{\delta}^\prime \mathbf{\Sigma}^{-1}\vec{\delta}$ between two populations. More precisely, we set $\vec{\mu}_1 =\mathbf{0}$ and the elements of $\vec{\mu}_2$ are randomly drawn from the uniform distribution
\begin{eqnarray*}
\left(\frac{e}{2},\frac{3e}{2}\right),\quad e=\frac{\Delta_L}{\beta},
\end{eqnarray*}
where
\begin{eqnarray*}
\Delta_L^2=(\mathbf{1}_{n_0}^\prime, \mathbf{0}_{p-n_0}^\prime)\mathbf{\Sigma}^{-1} (\mathbf{1}_{n_0}^\prime, \mathbf{0}_{p-n_0}^\prime)^\prime
\end{eqnarray*}
 is the Mahalanobis distance under the localized scenario, and $\beta$ is a parameter chosen to fulfill the requirement
\begin{eqnarray*}
\mathit{E}\Delta_D^2 =\mathit{E}\vec{\mu}_2^\prime \mathbf{\Sigma}^{-1}\vec{\mu}_2 =\Delta_L^2,
\end{eqnarray*}
where $\Delta_D^2$ is the Mahalanobis distance under the delocalized scenario. Direct calculations lead to
\begin{eqnarray*}
\beta^2= \frac{p (p\rho-14\rho+13)}{12(1-\rho+p\rho)(1-\rho)},
\end{eqnarray*}
for an equal correlation structure, $\mathbf{\Sigma}_{l,l^\prime}=\rho$ for $l\neq l^\prime$ and $\mathbf{\Sigma}_{ll}=1$.
For an autoregressive correlation structure, $\mathbf{\Sigma}_{l,l^\prime} = \rho^{|l-l^\prime|}$, we find
\begin{eqnarray*}
\beta^2=\frac{p(24\rho-13\rho^2-13)-24\rho+26\rho^2}{12(\rho^2-1)}.
\end{eqnarray*}
\end{enumerate}
By focusing on the delocalized scenario, simulation study is conducted to display the performances of proposed procedures.

As the main contribution of this paper, we generalize the D- and T- criteria from normality to general populations and establish their asymptotic misclassification probabilities. As it will be proven, the misclassification probability of the D-criterion will depend on the Mahalanobis distance between the two populations, and the misclassification probability of the T-criterion will depend on the difference of two group mean vectors and the skewness and kurtosis coefficients of the two populations $\Pi_1$ and $\Pi_2$.

The rest of the paper is organized as follows. In Section~\ref{sec:1}, the asymptotic misclassification probability of the D-criterion under general populations is derived and Monte Carlo experiments are conducted to compare the performance with that of several existing classification rules. In Section~\ref{sec:7},  the asymptotic misclassification probability of the T-criterion under general populations is derived. And a real data is used to present the competitive performance of the T-criterion. The conclusion is made at the end of the paper. Technical proofs are relegated to the appendix.
\section{The D-criterion}
\label{sec:1}
\subsection{Data generation model}
\label{sec:2}

Unlike the normal populations assumed in Saranadasa (1993), we assume that the populations $\Pi_1$ and $\Pi_2$ have the general form as introduced in Bai and Saranadasa (1996), i.e.

(a) The population $\mathbf{X}\sim \Pi_1$ has the form $\mathbf{X}=\Gamma \mathbf{X}^\ast +\vec{\mu}_1$, where $\Gamma$ is a $p\times p$ mixing or loading matrix, and $\mathbf{X}^\ast =(x_l^\ast)_{1\leq l \leq p}$ has $p$ independent and identically distributed, centered and standardized components. Moreover, $\gamma_x = \mathit{E}(|x_1^\ast|^4) < \infty$ and we set $\theta_x=\mathit{E}(x_1^{\ast 3})$.

(b) Similarly, the population $\mathbf{Y}\sim \Pi_2$ has the form $\mathbf{Y}=\Gamma \mathbf{Y}^\ast +\vec{\mu}_2$, where $\mathbf{Y}^\ast =(y_l^\ast)_{1\leq l \leq p}$ has $p$ independent and identically distributed, centered and standardized components. We set $\gamma_y =\mathit{E}(|y_1^\ast|^4) <\infty$ and $\theta_y =\mathit{E}(y_1^{\ast 3})$.

In consequence, the new observation $\mathbf{z} = \Gamma \mathbf{z}^\ast +\vec{\mu}_z$ where $\mathbf{z}^\ast = \mathbf{x}_i^\ast$ in distribution and $\vec{\mu}_z =\vec{\mu}_1$ if $\mathbf{z} \in \Pi_1$. Throughout the paper, we set $\tilde{\vec{\mu}} =\Gamma^{-1}\vec{\delta} = (\tilde{\mu})_{1\leq l \leq p}$, $\alpha_1=n_1/(n_1+1)$ and $\alpha_2=n_2/(n_2+1)$.

Notice that the data-generation model (a)--(b) are quite general meaning that the population are linear combinations of some unobservable independent component. They are also adopted in overall recent studies on high-dimensional statistics, see Chen et al. (2010), Li and Chen (2012), Srivastava et al. (2011) and etc.

\subsection{Asymptotic misclassification probability}
\label{sec:3}

The D-criterion (\ref{e1}) is easily seen equivalent to classifying $\mathbf{z}$ into $\Pi_1$ when
\begin{eqnarray}
\alpha_1 (\mathbf{z}-\bar{\mathbf{x}})^\prime \mathbf{A}^{-1}(\mathbf{z}-\bar{\mathbf{x}})<\alpha_2 (\mathbf{z} -\bar{\mathbf{y}})^\prime \mathbf{A}^{-1}(\mathbf{z}-\bar{\mathbf{y}}),\label{e3}
\end{eqnarray}
where
\begin{eqnarray}
\mathbf{A}=\sum_{i=1}^{n_1}(\mathbf{x}_i-\bar{\mathbf{x}})(\mathbf{x}_i -\bar{\mathbf{x}})^\prime + \sum_{j=1}^{n_2}(\mathbf{y}_j-\bar{\mathbf{y}})(\mathbf{y}_j -\bar{\mathbf{y}})^\prime,\label{e4}
\end{eqnarray}
involves correlation information between variables. This criterion has a straightforward form and does not need a preliminarily selected subset of features or any thresholding parameter.

The associated error of misclassifying $\mathbf{z}\in \Pi_1$ into $\Pi_2$ is
\begin{eqnarray}
P(2|1)=P\Big\{\alpha_1(\mathbf{z}-\bar{\mathbf{x}})^\prime \mathbf{A}^{-1}(\mathbf{z}-\bar{\mathbf{x}})-\alpha_2(\mathbf{z} -\bar{\mathbf{y}})^\prime \mathbf{A}^{-1}(\mathbf{z}-\bar{\mathbf{y}})>0 \big| \mathbf{z}\in \Pi_1\Big\}.\label{e5}
\end{eqnarray}
Under the data-generation models (a) and (b), since $\mathbf{x}_i =\Gamma \mathbf{x}_i^\ast +\vec{\mu}_1, \mathbf{y}_i =\Gamma \mathbf{y}_i^\ast +\vec{\mu}_2$, we have $\mathbf{A}=\Gamma \tilde{\mathbf{A}}\Gamma$, or $\Gamma \mathbf{A}^{-1}\Gamma =\tilde{\mathbf{A}}^{-1}$, where
\begin{eqnarray}
\tilde{\mathbf{A}} =\sum_{i=1}^{n_1}(\mathbf{x}_i^\ast- \bar{\mathbf{x}}^\ast)(\mathbf{x}_i^\ast- \bar{\mathbf{x}}^\ast)^\prime + \sum_{j=1}^{n_2} (\mathbf{y}_j^\ast -\bar{\mathbf{y}}^\ast) (\mathbf{y}_j^\ast-\bar{\mathbf{y}}^\ast)^\prime.\label{e6}
\end{eqnarray}
 The misclassification probability (\ref{e5}) is rewritten as
\begin{eqnarray}
P(2|1)&=&P\Big\{ \alpha_1(\mathbf{z}^\ast-\bar{\mathbf{x}}^\ast)^\prime \Gamma \mathbf{A}^{-1}\Gamma (\mathbf{z}^\ast-\bar{\mathbf{x}}^\ast) \nonumber\\
&&\hskip0.6cm - \alpha_2(\mathbf{z}^\ast-\bar{\mathbf{y}}^\ast
- \Gamma^{-1}\vec{\delta})^\prime \Gamma \mathbf{A}^{-1}\Gamma (\mathbf{z}^\ast-\bar{\mathbf{y}}^\ast - \Gamma^{-1}\vec{\delta}) >0 \big| \mathbf{z}\in \Pi_1\Big\} \nonumber\\
&=& P\Big\{\alpha_1(\mathbf{z}^\ast-\bar{\mathbf{x}}^\ast)^\prime \tilde{\mathbf{A}}^{-1}(\mathbf{z}^\ast-\bar{\mathbf{x}}^\ast) \nonumber\\
&&\hskip0.6cm -\alpha_2(\mathbf{z}^\ast -\bar{\mathbf{y}}^\ast -\tilde{\vec{\mu}})^\prime \tilde{\mathbf{A}}^{-1}(\mathbf{z}^\ast -\bar{\mathbf{y}}^\ast-\tilde{\vec{\mu}})>0 \big| \mathbf{z}\in \Pi_1\Big\}.\label{e7}
\end{eqnarray}

Here is the first main result of this paper.
\begin{theorem}\label{T1}
Under the data-generation models (a) and (b), assume that the following hold:
\begin{enumerate}
\item $p/n \to y \in (0,1)$ and $n_1/n \to \lambda \in (0,1)$, where $n=n_1+n_2-2$;

\item $\mathit{E}|x_{1}^\ast|^{4+b^\prime} < \infty$ and $\mathit{E}|y_{1}^\ast|^{4+b^\prime} < \infty$ for some constant $b^\prime >0$.
\end{enumerate}
Then as $p, n\to \infty$,  the misclassification probability (\ref{e7}) for the D-criterion satisfies
\begin{eqnarray}
\lim\left\{ P(2|1) -\Phi(\vartheta_1)\right\}=0,\label{e8}
\end{eqnarray}
where
\begin{eqnarray*}
\vartheta_1= -\frac{\Delta^2 }{2\sqrt{\frac{y}{\lambda (1-\lambda)} + \Delta^2}}\sqrt{1-y}, \quad
\Delta^2=||\tilde{\vec{\mu}}||^2=\vec{\delta}^\prime \mathbf{\Sigma}^{-1}\vec{\delta},
\end{eqnarray*}
is the Mahalanobis distance between the two populations $\Pi_1$ and $\Pi_2$.
\end{theorem}

The proof of the theorem is given in Appendix 1. The significance of the result is as follows. The asymptotic value of $P(2|1)$ depends on the values of $y$, $\lambda$ and $\Delta^2$, and is independent of other characteristics of the distributions $\Pi_1$ and $\Pi_2$. Firstly, this asymptotic value is symmetric about $\lambda$, so the value remains unchanged under a switch of the populations $\Pi_1$ and $\Pi_2$. Secondly, if $n_1$ and $n_2$ do not have large difference, i.e. $\lambda \rightarrow 0$ or $\lambda \rightarrow 1$, the asymptotic value of $P(2|1)$ mainly depends on $\Delta$ when $y$ is fixed. In other words, the classification task becomes easier for the D-criterion when the Mahalanobis distance between two populations increases as expected. However, when $y\to 1$, the number of features is very close to the sample size, the classification task becomes harder for the D-criterion due to the instability of the inverse $\mathbf{A}^{-1}$, a phenomenon well-noticed in high-dimensional statistical literature.

Under normal assumption, Saranadasa (1993) derived another asymptotic value for $P(2|1)$
\begin{eqnarray*}
\lim \left\{ P(2|1)-\Phi(\vartheta_2)\right)=0, \quad \vartheta_2= -\frac{1}{2}\Delta \sqrt{1-y}.
\end{eqnarray*}
Notice that $\vartheta_1=\tau\cdot \vartheta_2$, with
\begin{eqnarray*}
\tau=\frac{1}{\sqrt{\frac{y}{\lambda(1 -\lambda)\Delta^2} +1}}.
\end{eqnarray*}

\begin{figure}
\includegraphics[width=0.7\columnwidth]{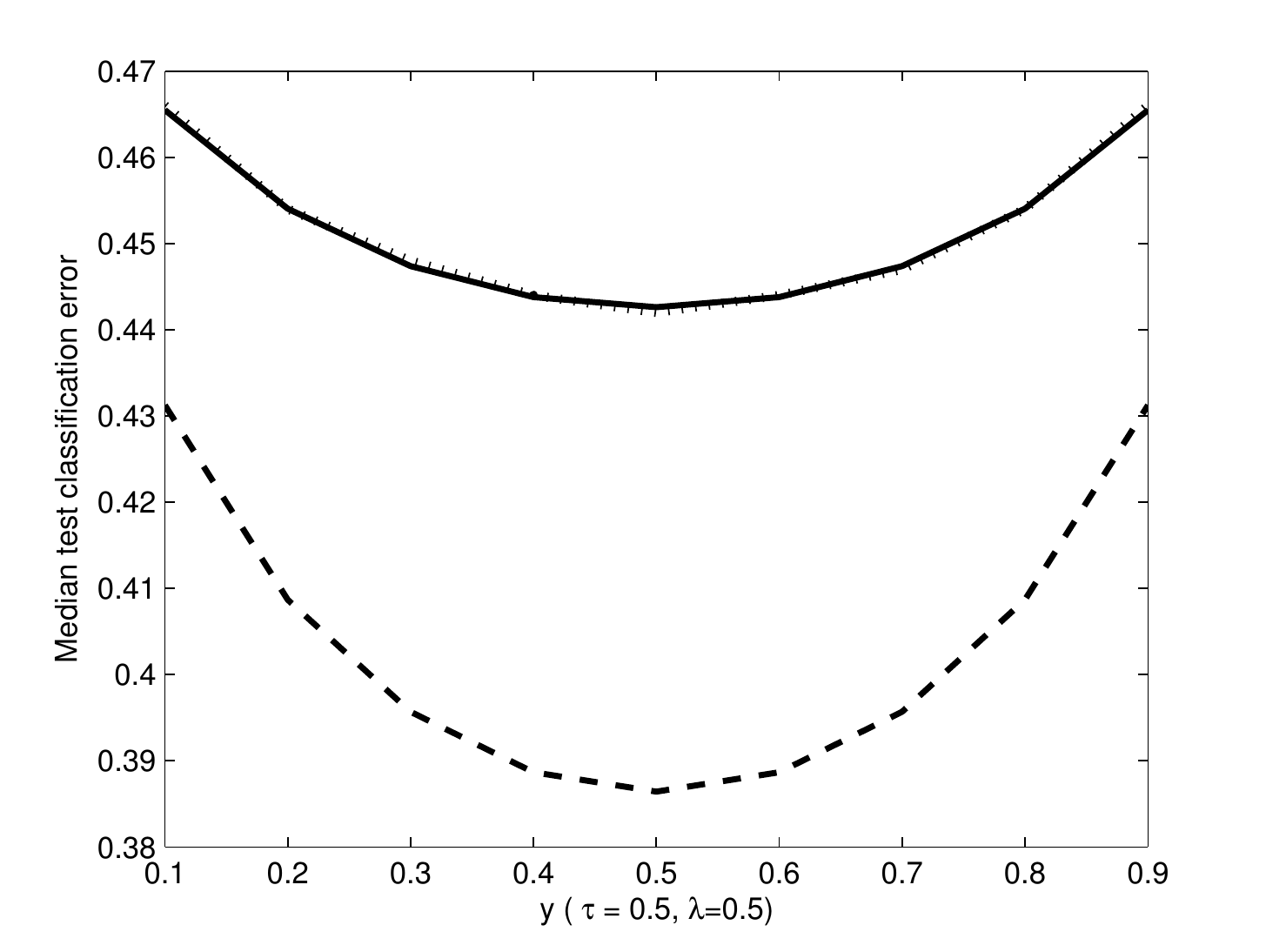}
\caption{Comparison between $\Phi(\vartheta_1)$ (solid), $\Phi(\vartheta_2)$ (dashes) and empirical values (dots) with 10,000 replications under normal samples. $n_1=n_2=500$ and $p$ ranges from 50 to 450 with step 50.}%
\label{fig:1}
\end{figure}

Let us comment on the difference between $\Phi(\vartheta_1)$ and $\Phi(\vartheta_2)$. The value of $\lambda$ does not influence on the difference significantly. Without loss of generality, let $\lambda=1/2$. The factor $\tau$ is 1/2 when $y$ and $\Delta^2$ satisfy $y/\Delta^2=3/4$. Under this setting, Figure~\ref{fig:1} shows the asymptotic values $\Phi(\vartheta_1)$, $\Phi(\vartheta_2)$ and compares them to empirical values from simulations, as $y$ ranges from 0.1 to 0.9 with step 0.1. Obviously, the difference between the two values are non-negligible ranging from $3.5\%$ to $5.5\%$. Moreover, $\Phi(\vartheta_1)$ is much closer to the empirical values than $\Phi(\vartheta_2)$. So our asymptotic result is more accurate. Other experiments have shown that only when the ratio of $y$ and $\Delta^2$ reaches some small values as of order $10^{-2}$, the difference between them can be negligible.

\begin{figure}%
\subfigure[]{%
\label{fig:2a}%
\includegraphics[width=0.52\columnwidth]{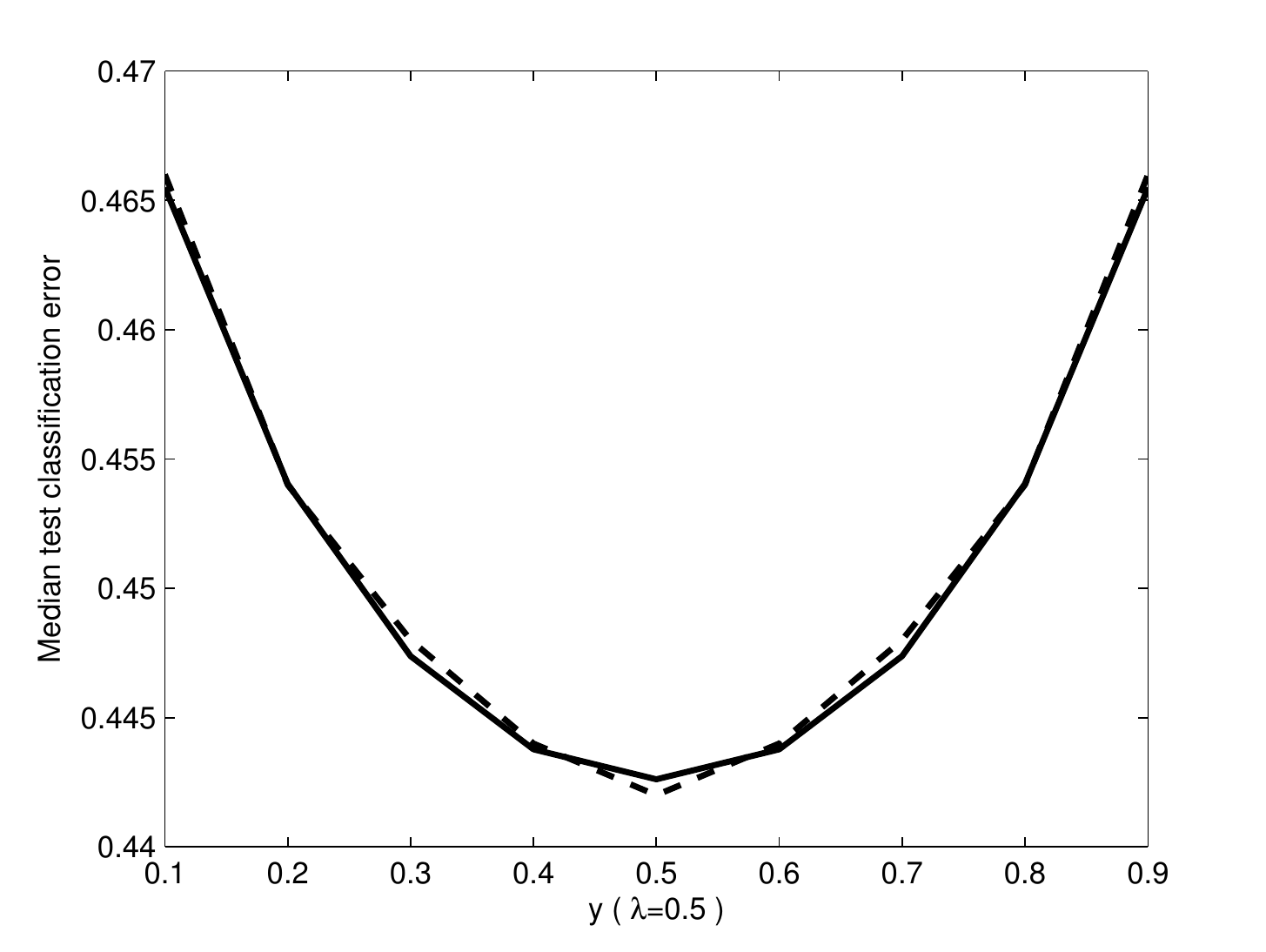}}%
\subfigure[]{%
\label{fig:2b} %
\includegraphics[width=0.52\columnwidth]{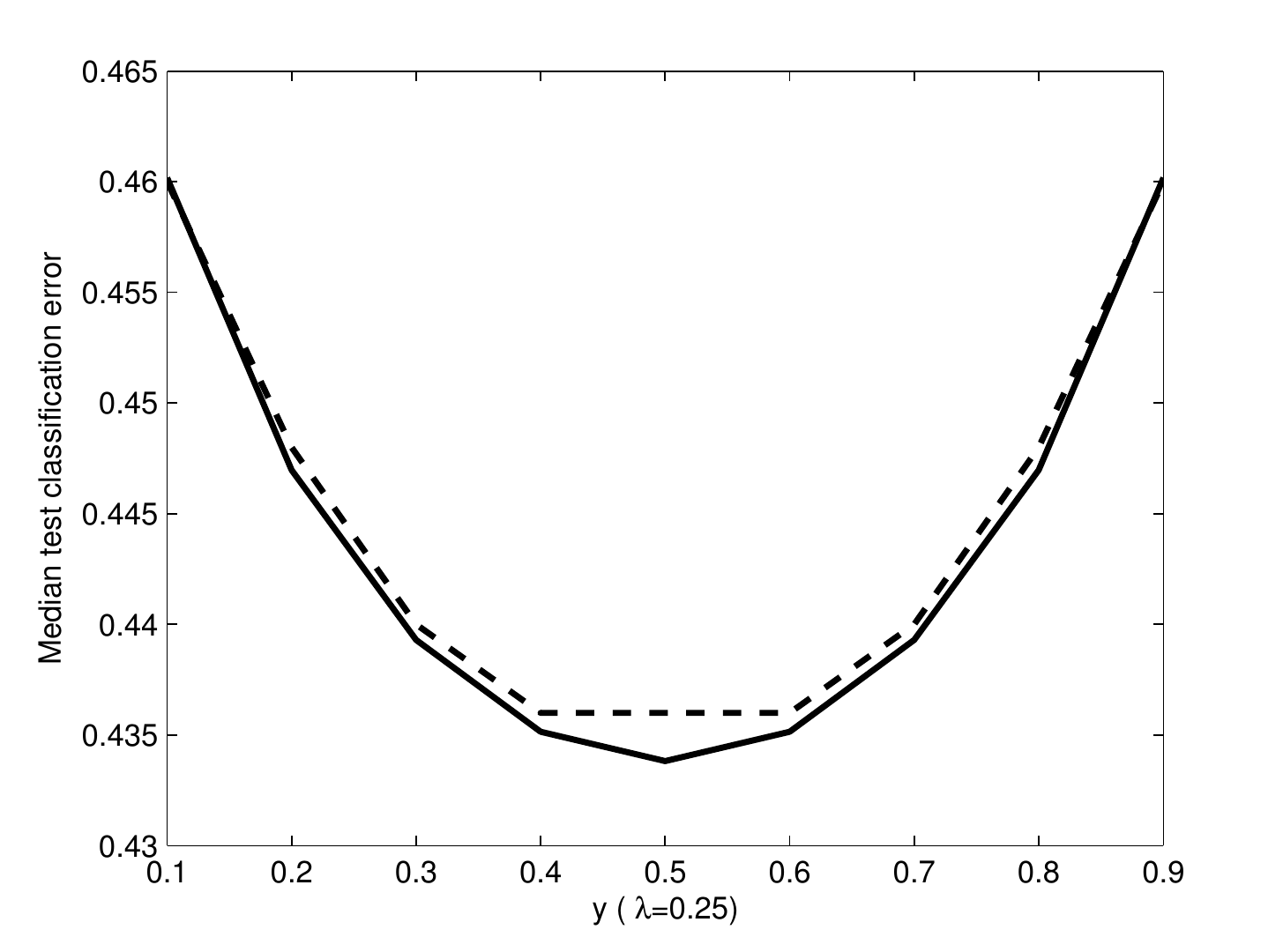}}
\caption{$\Phi(\vartheta_1)$ (solid) comparison of with empirical values (dashes) under different scenarios and with 10,000 replications for normal samples. Sample size $n=500$, and $n_1=n_2=n/2$ for the left, and $n_1=n/4$, $n_2=3n/4$ for the right. }%
\label{fig:2}%
\end{figure}

Additional experiments are conducted to check the accuracy of the asymptotic value $\Phi(\vartheta_1)$. Figure~\ref{fig:2} compares the values of $\Phi(\vartheta_1)$ to empirical values from simulations for normal samples. The empirical misclassification probabilities are very close to the theoretical values of $\Phi(\vartheta_1)$. It's the same for both $n_1=n_2$ and $n_1/n_2=1/4$ situations.

\subsection{Monte Carlo experiments}
\label{sec:4}
We conduct extensive tests to compare the D-criterion with several existing classification methods for high-dimensional data, the ROAD method and its variants S-ROAD1 andS-ROAD2, SCRDA, and the NB method, as well as the oracle. The oracle is defined following Fan et al. (2012) as the Fisher's LDA with true mean and true covariance matrix.

In all simulation studies, the number of variables is $p=125$, and the sample sizes of the training and testing data in two groups are $n_1=n_2=250$. The sparsity size is set to be $n_0=10$. A similar setting is used in Fan et al. (2012). Delocalized scenario is considered.

\subsubsection{Equal correlation setting}
\label{sec:5}

In this part, the covariance $\mathbf{\Sigma}$ is set to be an equal correlation matrix and correlation coefficient $\rho$ ranges from 0 to 0.9 with step 0.1.

\begin{figure}%
\centering
\subfigure[Median classification errors]{%
\label{fig:3a}%
\includegraphics[width=0.52\columnwidth]{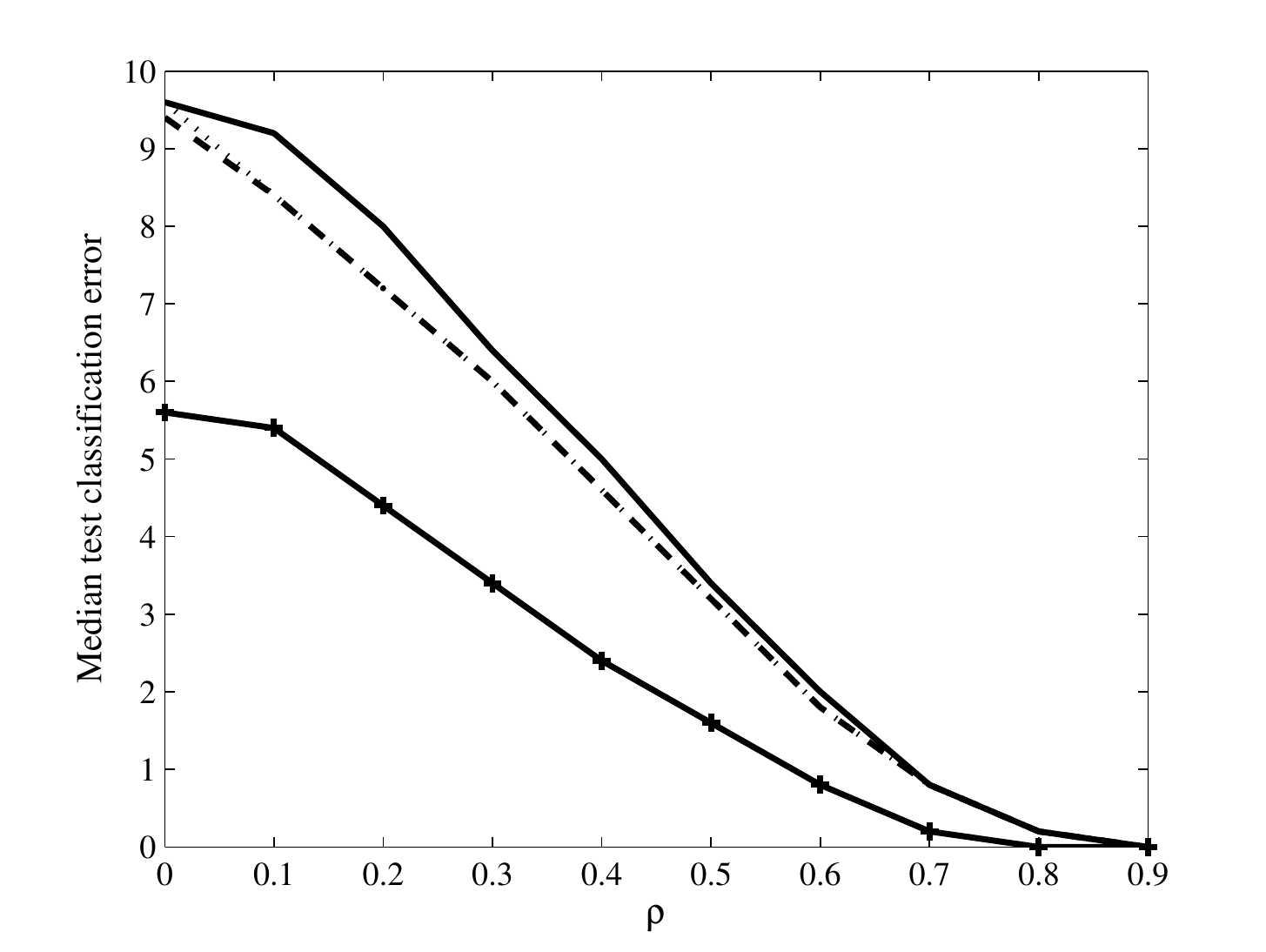}}%
\subfigure[Standard errors]{%
\label{fig:3b}%
\includegraphics[width=0.52\columnwidth]{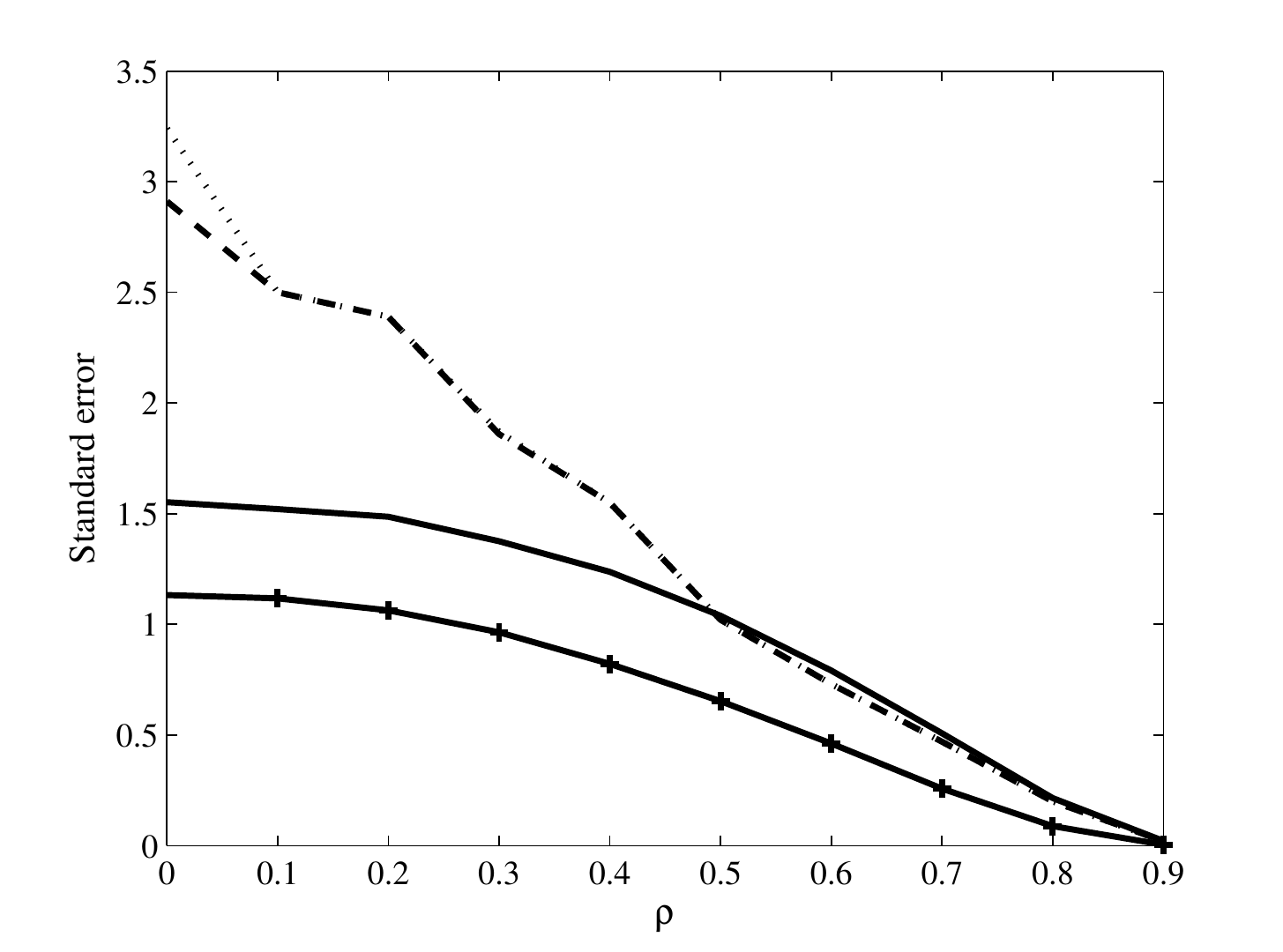}}%
\hspace{8pt}%
\caption{The median classification errors and standard errors for various methods under equal correlation structure and delocalization: D-criterion (solid); ROAD (dash); S-ROAD2 (dot); Oracle (cross).}%
\label{fig:3}%
\end{figure}

Simulation results for normal samples are shown in Table~\ref{tab:1} and a graphical summary is given in Figure~\ref{fig:3} including the median classification errors and standard errors.
The D-criterion performs similarly to the ROAD in terms of classification errors and is more robust than ROAD when $\rho$ is smaller than 0.5. The NB and the T-criterion lose efficiency when correlation exists in this setting. Notice that the results of SCRDA calculated using the R package provided by Guo et al. (2005) are not included. The package turns out to fail in some of our settings and report $``$NA" value. The percentage of failures in the simulations can reach 58\%. Therefore, it is unreliable to include SCRDA for comparison.

\begin{table}
 \caption{Comparison of the D-criterion with existing classifiers under the equal correlation setting for normal samples: median of test classification error (with their standard errors in parentheses)}
\label{tab:1}
\setlength{\tabcolsep}{3.5pt}
  \begin{tabular}{c|cccccccc}
\hline\noalign{\smallskip}
  $\rho$ & D-Criterion & ROAD & S-ROAD1 & S-ROAD2 & NB & Oracle & T-Criterion\\
\noalign{\smallskip}\hline\noalign{\smallskip}
  0 & 9.6(1.55)  & 9.4(2.91)&11.4(3.54)&9.6(3.24) &6.6(1.23)& 5.6(1.13) &6.2(1.18)\\
  0.1 & 9.2(1.52) &8.4(2.50)&8.6(2.58)&8.4(2.50) &12.4(1.57)&5.4(1.12) &12.4(1.57)\\
  0.2 & 8.0(1.49) &7.2(2.39)&7.4(2.42)&7.2(2.39) &16.8(1.77)&4.4(1.06) &16.8(1.76)\\
  0.3 & 6.4(1.37) &6.0(1.87)&6.0(1.86)&6.0(1.87) &20.2(1.88)&3.4(0.96) &20.2(1.87) \\
  0.4 & 5.0(1.24) &4.6(1.55)&4.6(1.55)&4.6(1.55) &22.6(1.94)&2.4(0.82) &22.6(1.94) \\
  0.5 & 3.4(1.04) &3.2(1.02)&3.2(1.02)&3.2(1.02) &24.6(2.00)&1.6(0.65) &24.6(1.99) \\
  0.6 & 2.0(0.79) &1.8(0.73)&1.8(0.74)&1.8(0.73) &26.2(2.04)&0.8(0.46) &26.2(2.03) \\
  0.7 & 0.8(0.51) &0.8(0.47)&0.8(0.47)&0.8(0.47) &27.4(2.06)&0.2(0.26) &27.4(2.05) \\
  0.8 & 0.2(0.22) &0.2(0.20)&0.2(0.20)&0.2(0.20) &28.6(2.08)&0.0(0.09) &28.6(2.07) \\
  0.9 & 0.0(0.02) &0.0(0.02)&0.0(0.02)&0.0(0.02) &29.6(2.10)&0.0(0.00) &29.6(2.10) \\
 \noalign{\smallskip}\hline
 \end{tabular}
\end{table}

\begin{table}
 \caption{Comparison of the D-criterion with existing classifiers under the equal correlation setting for Student's t samples: median of test classification error (with their standard errors in parentheses)}
 \label{tab:2}
\setlength{\tabcolsep}{3.5pt}
  \begin{tabular}{c|cccccccc}
\hline\noalign{\smallskip}
  $\rho$ & D-Criterion & ROAD & S-ROAD1 & S-ROAD2 & NB & Oracle & T-Criterion\\
\noalign{\smallskip}\hline\noalign{\smallskip}
  0 & 12.0(1.55)  & 9.0(2.76)&9.0(2.80)&9.0(3.24)
  &9.1(1.29)& 7.8(1.29) &8.6(1.24)\\
  0.1 & 11.6(1.56) &9.8(3.11)&15.2(6.32)&11.6(3.61) &15.2(4.17)&7.6(1.27) &14.8(3.40)\\
  0.2 & 10.4(1.48) &8.6(2.81)&19.6(6.76)&11.4(3.44) &19.2(7.00)&6.6(1.23) &19.0(5.79)\\
  0.3 & 9.0(1.38) &7.4(2.36)&24.0(7.26)&10.6(3.00) &22.4(8.83)&5.6(1.16) &22.0(7.58) \\
  0.4 & 7.6(1.27) &6.0(1.50)&27.6(8.06)&9.2(2.73) &24.8(10.15)&4.6(1.06) &24.2(8.99) \\
  0.5 & 6.0(1.13) &4.8(1.00)&28.9(9.35)&7.8(2.26) &27.0(11.11)&3.4(0.91) &26.2(10.11) \\
  0.6 & 4.4(0.97) &3.4(0.84)&29.2(10.83)&6.0(1.73) &29.0(11.90)&2.4(0.75) &27.6(11.02) \\
  0.7 & 2.8(0.78) &2.0(0.65)&29.2(12.32)&4.0(1.26) &30.6(12.51)&1.4(0.57) &29.0(11.79) \\
  0.8 & 1.2(0.53) &0.8(0.43)&28.8(13.74)&2.0(0.90) &32.0(13.01)&0.6(0.36) &30.2(12.44) \\
  0.9 & 0.2(0.23) &0.2(0.20)&28.6(15.06)&0.4(0.39) &33.4(13.35)&0.0(0.14) &31.2(12.96) \\
\noalign{\smallskip}\hline
 \end{tabular}
\end{table}
Simulation results for Student's t (degree of freedom is set to be 7) samples are shown in Table~\ref{tab:2}. All classifiers have slightly higher misclassification rates for Student's t samples. S-ROAD1 and S-ROAD2 have larger standard errors. And S-ROAD1, NB and T-criterion lose efficiency when correlation is significant. The D-criterion outperforms the others except ROAD in term of classification error. But the D-criterion has the smallest standard error which is close to that of Oracle.

\subsubsection{Autoregressive correlation setting}
\label{sec:6}
In this part, the covariance $\mathbf{\Sigma}$ is set to be an autoregressive correlation matrix and $\rho$ ranges from 0 to 0.9 with step 0.1. Previous the results have shown that NB is not a good rule when significant correlation exists. Therefore, NB is no more included in comparison. Since the comparison results are similar in normal samples and Student't t samples, we only use normal samples in this part.

\begin{figure}%
\subfigure[Median classification errors]{%
\label{fig:4a}%
\includegraphics[width=0.52\columnwidth]{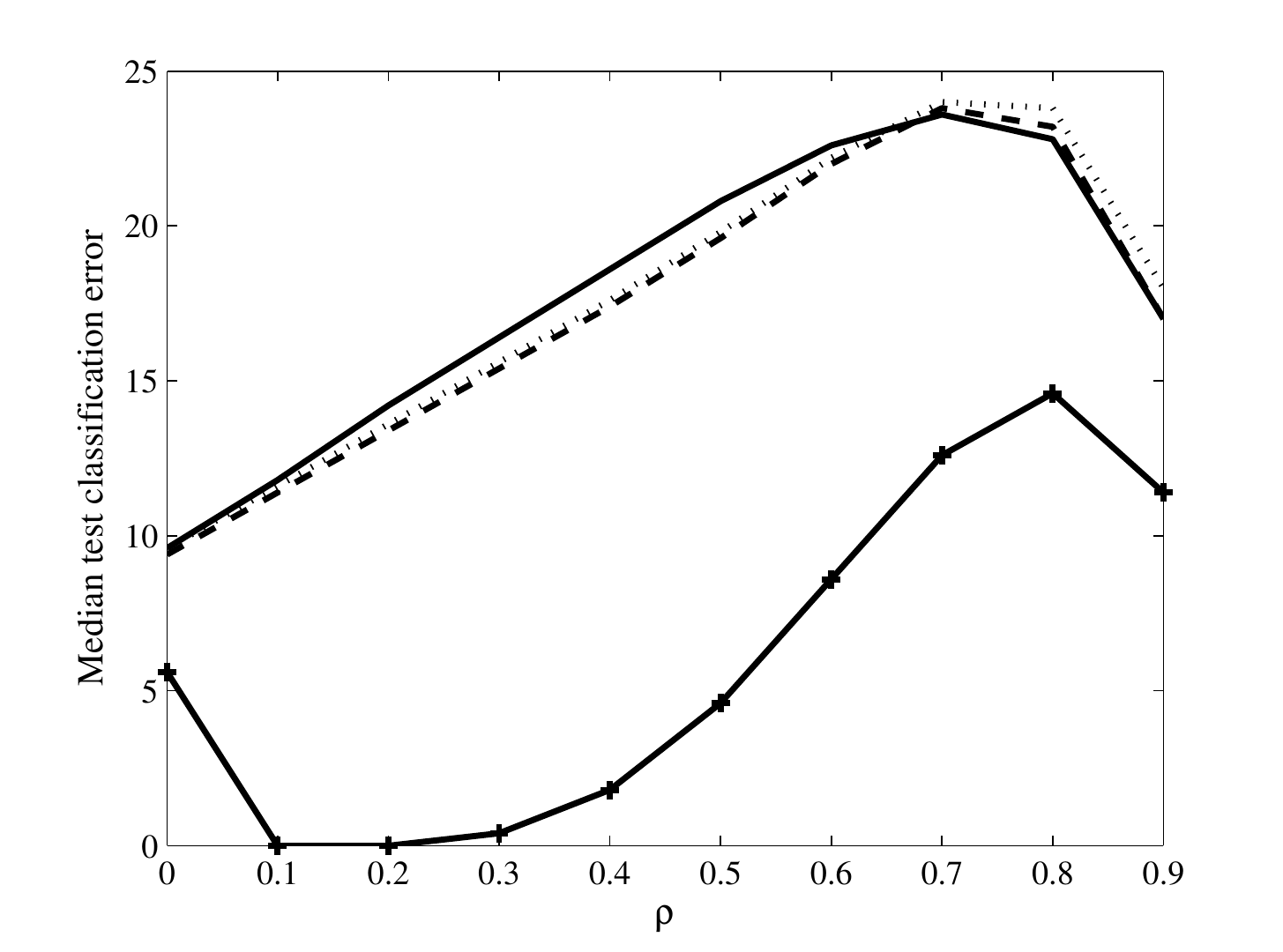}}%
\subfigure[Standard errors]{%
\label{fig:4b}%
\includegraphics[width=0.52\columnwidth]{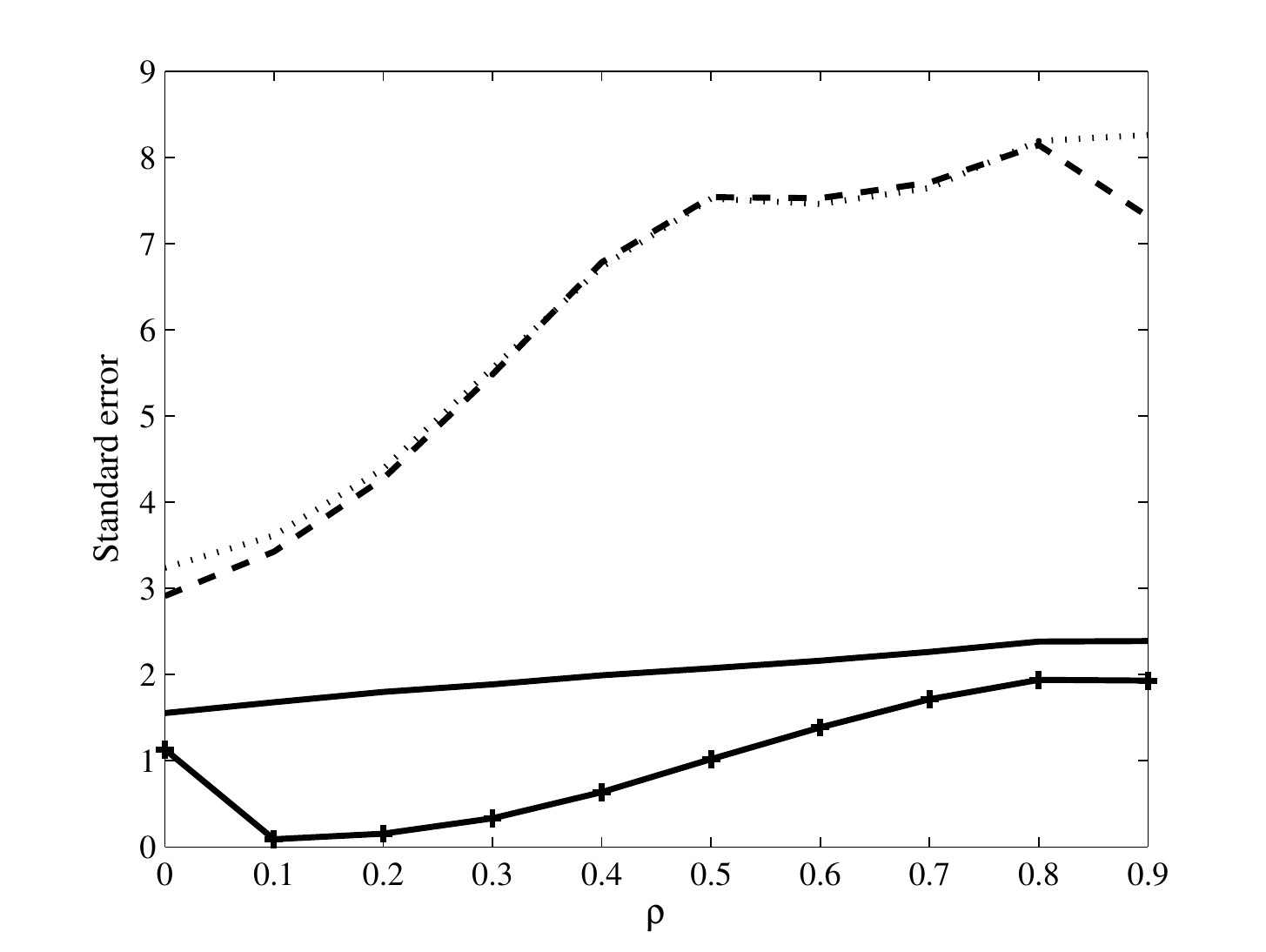}}%
\hspace{8pt}%
\caption{The median classification errors and standard errors for various methods under autoregressive correlation structure: D-criterion (solid); ROAD (dash); S-ROAD2 (dot); Oracle (cross).}%
\label{fig:4}%
\end{figure}

Simulation results are shown in Table~\ref{tab:3} and a graphical summary is given in Figure~\ref{fig:4}. The T-criterion is only suitable for independent case $\rho=0$, and loses efficiency when $\rho>0$. The D-criterion has the same performance with ROAD and S-ROAD2 in terms of classification error. Moreover, the D-criterion is much more robust and has a standard error close to that of the oracle.

\begin{table}
\caption{Comparison of the D-criterion with existing classifiers under the autoregressive correlation setting: median of test classification error (with their standard errors in parentheses)}
\label{tab:3}
\begin{tabular}{c|ccccccc}
\hline\noalign{\smallskip}
$\rho$ & D-Criterion & ROAD & S-ROAD1 & S-ROAD2  & Oracle & T-Criterion\\
\noalign{\smallskip}\hline\noalign{\smallskip}
0 & 9.6 (1.55) & 9.4 (2.91)& 11.6 (3.54)&9.6 (3.24) & 5.6(1.13)&6.2(1.18)\\
0.1 &11.8(1.68) &11.4(3.42)&12.8(3.67)&11.6(3.61)&
0.0(0.09)&8.0(1.31)\\
0.2 &14.2(1.80) &13.4(4.27)&14.4(4.02)&13.6(4.39)&0.0(0.15)&10.0(1.44)\\
0.3 &16.4(1.89) &15.4(5.48)&16.0(4.61)&15.6(5.55)&0.4(0.33)&12.2(1.57)\\
0.4 &18.6(1.99) & 17.4(6.78)&17.8(5.95)&17.6(6.73)&1.8(0.64)&14.8(1.70)\\
0.5 &20.8(2.07) &19.6(7.54)&20.0(7.29)&19.8(7.52)&4.6(1.02)&17.8(1.81)\\
0.6 &22.6(2.16) &22.0(7.53)&22.6(7.34)&22.2(7.46)&8.6(1.38)&21.4(1.92)\\
0.7 &23.6(2.26) &23.8(7.71)&26.0(7.54)&24.0(7.64)&12.6(1.71)&25.0(2.03)\\
0.8 &22.8(2.38) &23.2(8.14)&30.6(7.67)&23.8(8.19)&14.6(1.94)&31.0(2.12)\\
0.9 &17.0(2.39) &17.0(7.31)&33.4(9.13) &18.0 (8.26)&11.4(1.93)&37.0(2.19)\\
\noalign{\smallskip}\hline
 \end{tabular}
\end{table}

In conclusion, compared to these existing methods, the D-criterion is competitive for $``$large $p$-large $n$" situation specifically under delocalized scenario and autoregressive correlation structure. In such a scenario, the D-criterion has a classification error comparable to that of the Road-family classifiers while being the most robust with a much smaller standard error close to that of the oracle.

\section{The T-criterion}
\label{sec:7}
Notice that one limitation of the D-criterion is that the dimension $p$ must be smaller than the sample size $n$. In addition, when the ratio $p/n$ is close to $1$, the performance of this criterion becomes bad due to the matrix $\mathbf{A}$ is close to singular. The T-criterion in contrast does not have such a limitation.

\subsection{Asymptotic misclassification probability}
\label{sec:8}
The T-criterion (\ref{e2}) is easily seen equivalent to
\begin{eqnarray}
\alpha_1 (\mathbf{z}-\bar{\mathbf{x}})^\prime (\mathbf{z}-\bar{\mathbf{x}}) < \alpha_2 (\mathbf{z}-\bar{\mathbf{y}})^\prime (\mathbf{z}-\bar{\mathbf{y}}).\label{e9}
\end{eqnarray}
Obviously, the T-criterion has a very simple form only involving the group mean vectors. In particular, it does not require to select a subset of features or to choose a threshold parameter.

When $\mathbf{z}\in \Pi_1$, the error of misclassifying $\mathbf{z}$ into $\Pi_2$ is
\begin{eqnarray}
P(2|1)=P\Big\{\alpha_1(\mathbf{z} -\bar{\mathbf{x}})^\prime(\mathbf{z} -\bar{\mathbf{x}}) -\alpha_2(\mathbf{z}-\bar{\mathbf{y}})^\prime(\mathbf{z}-\bar{\mathbf{y}})>0 \big| \mathbf{z}\in \Pi_1 \Big\}.\label{e10}
\end{eqnarray}
Here is the second main result of this paper. Throughout the paper, $\mathbf{1}_d$ is a length $d$ vector with all entries 1, $\mathbf{0}_d$ is a length $d$ vector with all entries 0.

\begin{theorem}\label{T2}
Under the data-generation models (a) and (b), assume that the following hold:
\begin{enumerate}
\item $\gamma_{4+b^\prime, x} =\mathit{E}|x_{1}^\ast|^{4+b^\prime} < \infty$ and $\gamma_{4+b^\prime, y} =\mathit{E}|y_{1}^\ast|^{4+b^\prime} < \infty$ for some constant $b^\prime >0$;

\item the covariance matrix $\mathbf{\Sigma}$ is diagonal, i.e. $\vec{\Sigma} = \textrm{diag} (\sigma_{ll})_{1\leq l \leq p}$;

\item $\sup_{p\geq 1} \left\{ |\delta_l|, \sigma_{ll}\delta_l^2, l=1, \ldots, p\right\} <\infty$; and

\item $\displaystyle \frac{  \sum_{l=1}^p \sigma_{ll}^{2+b} +\sum_{l=1}^p \delta_l^{4+2b}}{ \left(\sum_{l=1}^p \sigma_{ll}\delta_l^2\right)^{1+\frac{b}{2}}}\to 0$ as $p \to \infty$, where $b=b^\prime/2$.
\end{enumerate}
Then we have as $p\to \infty$ and $n_\ast =\min (n_1, n_2) \to \infty$,
\begin{eqnarray}
\lim \left\{ P(2|1) - \Phi \left(  -\frac{\alpha_2 ||\vec{\delta}||^2}{B_p} \right) \right\} = 0,\label{e11}
\end{eqnarray}
where
\begin{eqnarray*}
B^2_p &=& 4\left(\frac{1}{n_1}+\frac{1}{n_2}\right)\textrm{tr}(\mathbf{\Sigma}^2) + 4\theta_x \left(\frac{1}{n_2}-\frac{1}{n_1}\right)\mathbf{1}_p^\prime \Gamma^3 \vec{\delta}  \\ &&+4\left(1-\frac{1}{n_2}\right)\vec{\delta}^\prime \mathbf{\Sigma}\vec{\delta} + O\left(\frac{p}{n_\ast^2}\right).
\end{eqnarray*}
\end{theorem}
The proof of the theorem is given in Appendix 2. Assumption 1 is needed for dealing with non-normal populations. Assumption 3 is a weak and technical condition without any practical limitation. Assumption 4 is satisfied for most applications where typically $\sum_{l=1}^p \sigma_{ll}^{2+b}, \sum_{l=1}^p \sigma_{ll}\delta_l^2$ and $\sum_{l=1}^p \delta_l^{4+2b} $ are all of order $p$. The main term of $B_p^2$ is,
\begin{eqnarray*}
 B_p^2 \approx 4 \vec{\delta}^\prime \mathbf{\Sigma}\vec{\delta},
 \end{eqnarray*}
since it has the order $O(p)$ and other terms are $O(p/n_\ast)$. In order to get more accurate result in finite sample case, these $O(p/n_\ast)$ terms are kept in the Theorem.

Notice that the main term of the approximation of $P(2|1)$ depends on the ratio $(\vec{\delta}^\prime \vec{\delta})/(2\sqrt{\vec{\delta}^\prime \vec{\Sigma} \vec{\delta}})$. If the components $\delta_l$ of $\vec{\delta}$ satisfy $|\delta_l|\geq c >0$, and $0< d_1 \leq \lambda_{min}(\vec{\Sigma}) \leq \lambda_{max}(\vec{\Sigma}) \leq d_2$ for positive constants $c, d_1, d_2$, then when $p\to \infty$,
 \begin{eqnarray*}
 \vec{\delta}^\prime \mathbf{\Sigma}\vec{\delta}\geq p d_1 c^2 \to \infty,
 \end{eqnarray*}
 and
\begin{eqnarray*}
\frac{\vec{\delta}^\prime \vec{\delta}}{2\sqrt{\vec{\delta}^\prime \mathbf{\Sigma} \vec{\delta}}} \geq \frac{1}{2\sqrt{d_2}}\cdot \sqrt{\vec{\delta}^\prime \mathbf{\Sigma}\vec{\delta}}\to \infty, \quad \textrm{i.e.} \ P(2|1)\to 0.
\end{eqnarray*}
In other words, the classification task becomes easier when the dimension grows. In other scenarios, this misclassification probability is not guaranteed to vanish. For example, under a localized scenario, $\delta_1 = \cdots = \delta_{n_0} =c \neq 0$, $\delta_l =0$ for $l>n_0$ and $n_0$ is fixed and independent of $p$, then
\begin{eqnarray*}
\frac{c}{2}\sqrt{\frac{n_0}{d_2}}\leq\frac{\vec{\delta}^\prime \vec{\delta}}{2\sqrt{\vec{\delta}^\prime \mathbf{\Sigma} \vec{\delta}}}\leq \frac{c}{2}\sqrt{\frac{n_0}{d_1}}, \quad \textrm{i.e.} \ \lim \inf P(2|1)\geq \Phi\left( - \frac{c}{2}\sqrt{\frac{n_0}{d_1}} \right) to 0.
\end{eqnarray*}

\begin{figure}%
\centering
\subfigure[Normal samples]{%
\label{fig:a}%
\includegraphics[width=0.52\columnwidth]{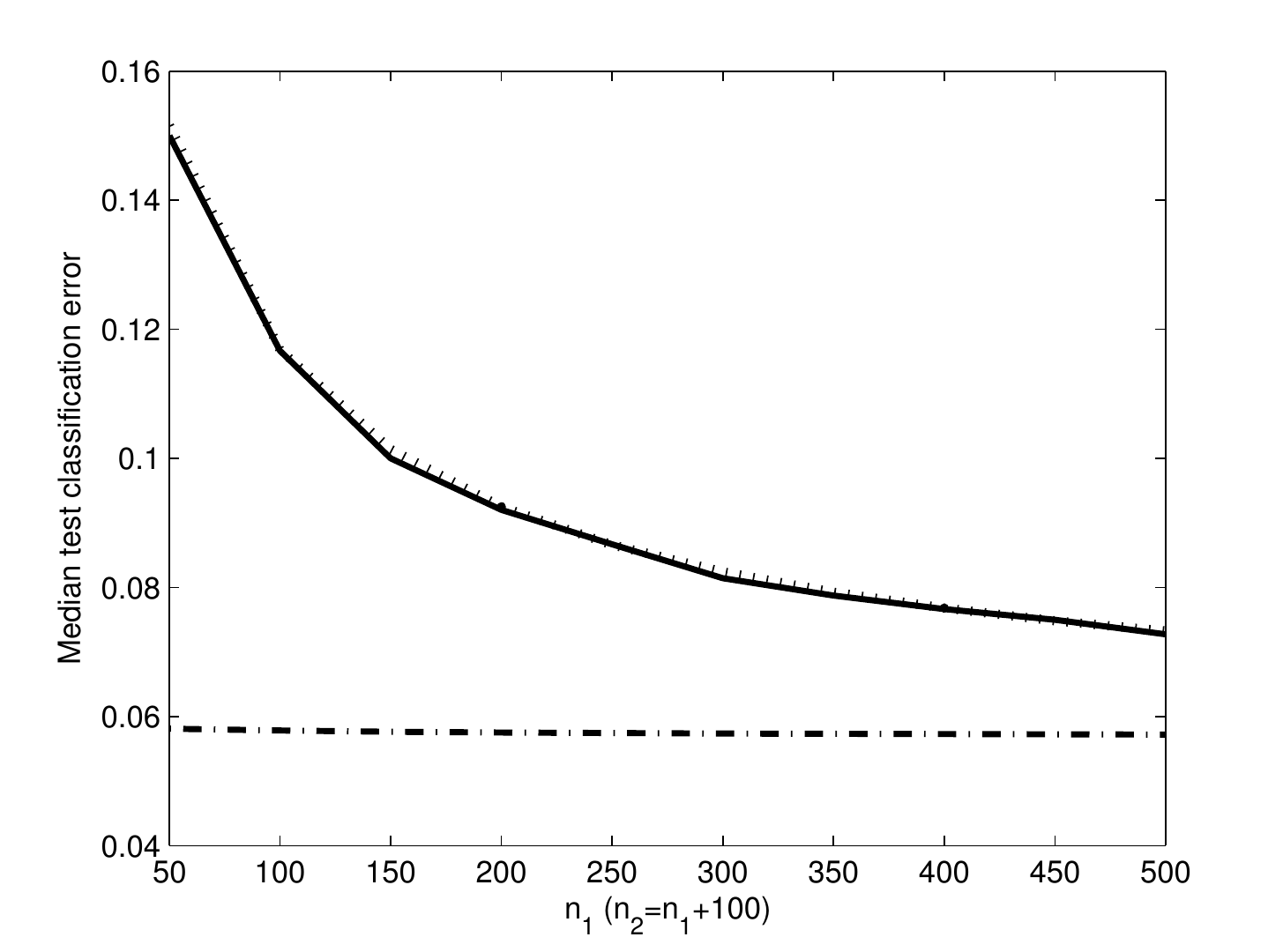}}%
\subfigure[Gamma samples]{%
\label{fig:b}%
\includegraphics[width=0.52\columnwidth]{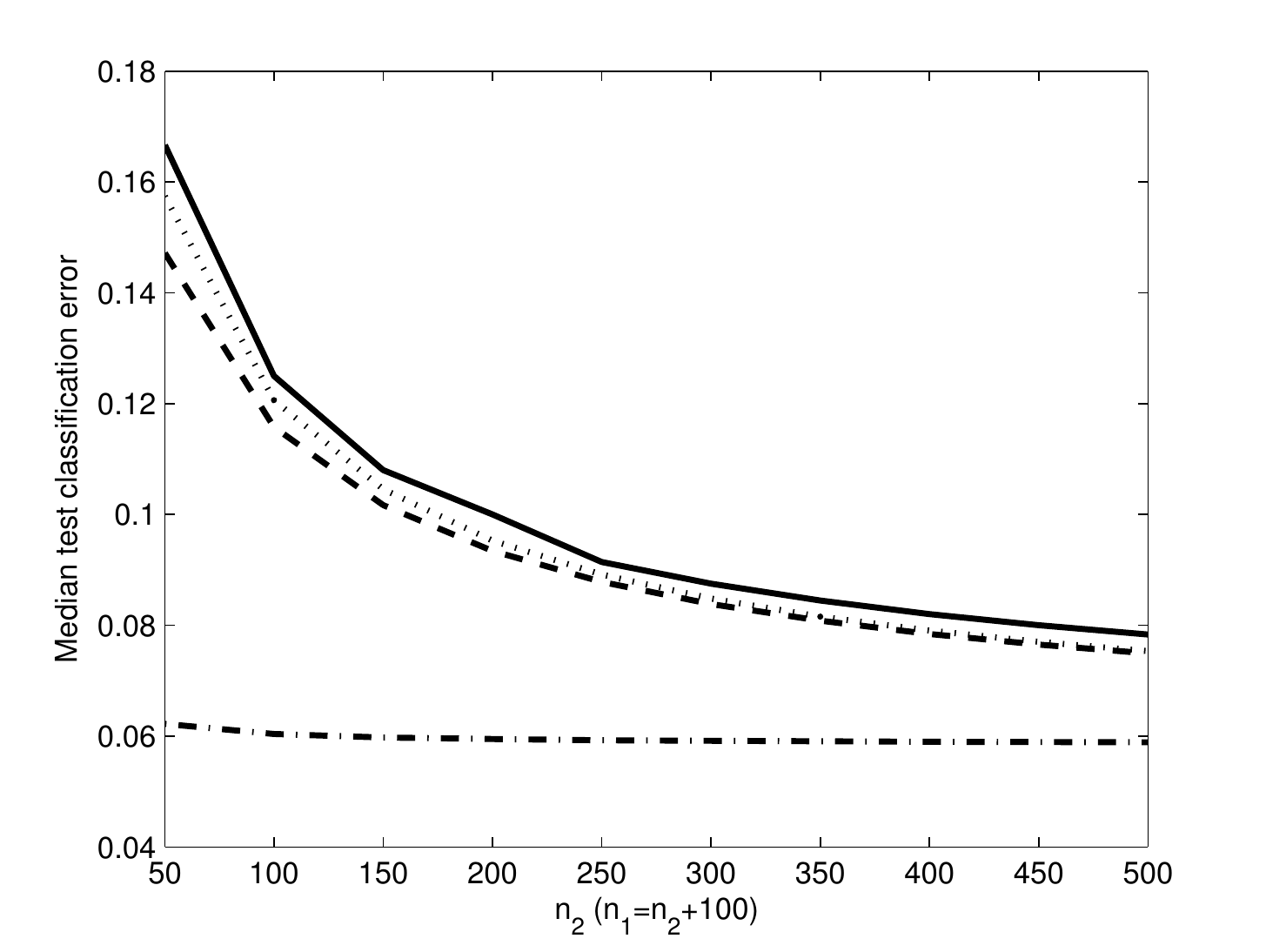}}%
\hspace{8pt}%
\caption{The empirical values (solid) are compared to asymptotic values: dots ($B_p^2(1)$), dashes ($B_p^2(2)$) and dash-dots($B_p^2(3)$), with 10,000 replications for normal samples and gamma samples. $p=500$, $n_1$ range from 50 to 500 with step 50 and $n_2=n_1+100$.}%
\label{fig:5}%
\end{figure}

Next, we provide below some simulation results to demonstrate the importance of keeping the $O(p/n_\ast)$ terms in $B_p^2$. The experiments use $p=500$ and various combinations of sample sizes $(n_1, n_2)$ with normal samples and gamma samples, respectively. Empirical classification errors are compared in Figure~\ref{fig:5} to the following three approximations of the variance $B_p^2$:
\begin{itemize}
\item $B_p^2(1)= 4\left(\frac{1}{n_1}+\frac{1}{n_2}\right)\textrm{tr}(\mathbf{\Sigma}^2) + 4\theta_x \left(\frac{1}{n_2}-\frac{1}{n_1}\right)\mathbf{1}_p^\prime \Gamma^3 \vec{\delta} +4\left(1-\frac{1}{n_2}\right)\vec{\delta}^\prime \mathbf{\Sigma}\vec{\delta}$;

\item $B_p^2(2)= 4\left(\frac{1}{n_1}+\frac{1}{n_2}\right)\textrm{tr}(\mathbf{\Sigma}^2) + 4\left(1-\frac{1}{n_2}\right)\vec{\delta}^\prime \mathbf{\Sigma}\vec{\delta} $;

\item $B_p^2(3)= 4\vec{\delta}^\prime \mathbf{\Sigma}\vec{\delta} $.
\end{itemize}
Among the three, the proposed approximation $B_p^2(1)$ matches very well the empirical values, while $B_p^2(3)$ is by far the worst in all tested cases.
As for $B_p^2(1)$ and $B_p^2(2)$, they are by definition the same for normal samples (since $\theta_x=0$). For gamma samples, they remain close each other particularly when the relative difference of sample sizes $(1/n_2 -1/n_1)$ become small, and $B_p^2(1)$ has an overall slightly better performance than $B_p^2(2)$ (in these tested cases). Notice that the gamma standardized variables are $\mathbf{x}_i^\ast=(\mathbf{u}_i-1)$ where $\mathbf{u}_i$ is gamma distributed with unit shape and scale parameters so that $\theta_x=-2$.

Under normal assumption, the expectation of $\sum_{l=1}^p k_l$ (defined in Appendix) is the same with (\ref{e18}), and the variance simplifies to
\begin{eqnarray*}
B_p^2= 4\left(\frac{1}{n_1}+\frac{1}{n_2}\right)\textrm{tr}(\mathbf{\Sigma}^2) +4\left(1-\frac{1}{n_2}\right)\vec{\delta}^\prime \mathbf{\Sigma}\vec{\delta} +O\left(\frac{p}{n_\ast^2}\right),
\end{eqnarray*}
which coincides with the result established in Saranadasa (1993).

\subsection{Monte Carlo experiments}
\label{sec:9}
We conduct simulations to show the performances of the T-criterion for normal distributions under delocalized scenario. In the simulation studies, the number of variables is $p=500$. Without loss of generality, the sample sizes of the training and testing data in two groups are equal and range from 100 to 500 with step 50. The covariance $\mathbf{\Sigma}$ is set to be an identity matrix $\mathbf{I}_p$ and the sparsity size is $n_0=10$.

Simulation results are shown in Table~\ref{tab:4}. The classification error decreases as sample size increases. Meanwhile, small standard errors indicate that the T-criterion is robust with respect to the delocalization nature of mean differences. Notice that the T-criterion is an independence rule. It's suitable for case where variables are independent or the correlations between variables are weak. As shown in Tables 1-3, the T-criterion has very high misclassification rate when variables have significant correlations.
\begin{table}
\caption{The T-criterion under delocalization setting: median of test classification errors (with standard errors in parentheses)}
\label{tab:4}
\setlength{\tabcolsep}{3.5pt}
\begin{tabular}{c|cccc|ccccc}
\hline\noalign{\smallskip}
\ & \multicolumn{4}{c}{$p>n$ }& \multicolumn{5}{c}{$p<n$}\\
\noalign{\smallskip}\hline\noalign{\smallskip}
$n_1=n_2$  & 100 & 150 & 200 & 250 & 300 & 350 & 400 & 450 & 500\\
\noalign{\smallskip}\hline\noalign{\smallskip}
median& 13.00 & 11.00 & 9.75 & 9.00 & 8.50 & 8.14 & 7.88 & 7.56 & 7.40 \\
s.e. & (2.52) & (1.90) & (1.57) & (1.35) & (1.20) & (1.11) & (1.01) & (0.95) & (0.89) \\
\noalign{\smallskip}\hline\noalign{\smallskip}
\end{tabular}
\end{table}

\subsection{A real data analysis}
\label{sec:10}
In this part, we analyze a popular gene expression data: $`$leukemia' (Golub et al. 1999). The leukemia data set contains $p=7129$ genes for $n_1=27$ acute lymphoblastic leukemia and $n_2=11$ acute myeloid leukemia vectors in the training set. The testing set includes 20 acute lymphoblastic leukaemia and 14 acute myeloid leukemia vectors. Obviously, this data set is a $``$large $p$-small $n$" case. The classification results for the T-criterion, ROAD, SCRDA, FALR, NSC and NB methods are shown in Table~\ref{tab:5}. (The results for ROAD, SCRDA, FAIR, NSC and NB are found in Fan et al. (2012).) The T-criterion is as good as ROAD and NB in terms of training error. ROAD and FAIR perform better than T-criterion in terms of testing error. Both of NB and T-criterion make use of all genes, but T-criterion outperforms NB. Meanwhile, T-criterion performs better than NSC. Overall, on this data set, T-criterion outperforms SCRDA, NSC and NB, equally well as FIRE, and is beaten only by ROAD (2 v.s. 1 errors). It's quite surprising that a $``$simple-minded$"$ rule like T-criterion has a performance comparable to a sophisticated rule like ROAD.
\begin{table}
\caption{Classification error and number of used genes for the leukemia data}
\label{tab:5}
\begin{tabular}{|c|c|c|c|}
\hline\noalign{\smallskip}
Method & Training error & Testing error & Number of genes used\\
\noalign{\smallskip}\hline\noalign{\smallskip}
T-criterion & 0 & 2 & 7129\\
ROAD & 0 & 1 & 40\\
SCRDA & 1 & 2 & 264\\
FAIR & 1 & 1 & 11\\
NSC & 1 & 3 & 24\\
NB & 0 & 5 & 7129\\
\noalign{\smallskip}\hline
\end{tabular}
\end{table}

\section{Conclusion}
\label{sec:11}
We have proposed two new classification rules for high-dimensional data, namely the D-criterion and the T-criterion. Both methods consider the overall within group sum of squares and cross products matrices. The D-criterion compares the
determinants of these matrices and integrates correlation information between variables. The D-criterion performs well
when correlations between variables become significant. When the correlation coefficient increases, the classification error
of the D-criterion drops. The incorporation of covariance structure therefore strengthens the effectiveness
in high dimensional classification.  The T-criterion, on the other hand,
compares the traces of these matrices and involves only group mean vectors.
The implementation of these two criteria is straightforward and it does
not suffer from challenging issues such as variable selection,
thresholding or control of the sparsity size that are required
in the existing methods. We found D-criterion is particularly
competitive in delocalized scenario. When $p>n$, the T-criterion
is quite effective as proven by the real data analysis.

Moreover, using the explicit forms of the criteria and recent results from random matrix theory, we are able to derive asymptotic approximations for the misclassification probability of both criteria. Notice that such asymptotic approximations are unknown for most of the existing high-dimensional classifiers in the literature. Simulation results have shown that the proposed approximations are quite accurate for both normal and non-normal populations.

\appendix

\section{Appendix Technical proofs}
\label{sec:12}

\subsection{Proof of Theorem 1}
\label{sec:13}
We first recall two known results on the Mar\v{c}enko-Pastur distribution, which can be found in Theorem 3.10 in Bai and Silverstein (2010) and Lemma 3.1 in Bai et al. (2009).
\begin{lemma}\label{L1}
Assume $p/n\to y\in (0,1)$ as $n\to \infty$, for the sample covariance matrix $\tilde{\mathbf{S}}= \tilde{\mathbf{A}}/n$, we have the following results
\begin{enumerate}
\item[(1)]
\begin{eqnarray*}
\frac{1}{p}tr(\tilde{\mathbf{S}}^{-1}) \stackrel{a.s.}{\longrightarrow} a_1, \quad \frac{1}{p}tr(\tilde{\mathbf{S}}^{-2}) \stackrel{a.s.}{\longrightarrow} a_2,
\end{eqnarray*}
where $a_1=\frac{1}{1-y}$ and $a_2=\frac{1}{(1-y)^3}$;

\item[(2)] Moreover,
\begin{eqnarray*}
\bar{\mathbf{x}}^{\ast\prime}\tilde{\mathbf{S}}^{-i} \bar{\mathbf{x}}^\ast \stackrel{a.s.}{\longrightarrow} a_i,\quad \bar{\mathbf{y}}^{\ast\prime}\tilde{\mathbf{S}}^{-i} \bar{\mathbf{y}}^\ast \stackrel{a.s.}{\longrightarrow} a_i, i=1, 2.
\end{eqnarray*}
\end{enumerate}
\end{lemma}

Under the data-generation models (a) and (b), let $
 \Omega =(\tilde{\mathbf{A}}, \bar{\mathbf{x}}^\ast, \bar{\mathbf{y}}^\ast)$. Conditioned on $\Omega$, the misclassification probability (\ref{e7}) can be rewritten as
\begin{eqnarray*}
P_{\Omega}(2|1)&=&P \left( K >0 \big| \Omega\right)=
P_{\Omega}\left(K >0\right),
\end{eqnarray*}
where
\begin{eqnarray*}
K= \alpha_1 (\mathbf{z}^\ast -\bar{\mathbf{x}}^\ast)^\prime \tilde{\mathbf{A}}^{-1} (\mathbf{z}^\ast -\bar{\mathbf{x}}^\ast) - \alpha_2 (\mathbf{z}^\ast -\bar{\mathbf{y}}^\ast - \tilde{\vec{\mu}})^\prime \tilde{\mathbf{A}}^{-1} (\mathbf{z}^\ast -\bar{\mathbf{y}}^\ast - \tilde{\vec{\mu}}).
\end{eqnarray*}
Therefore, $\displaystyle P_\Omega(2|1) =P_\Omega \left( K >0 \right)$ where $\mathbf{z}\in \Pi_1$ is assumed implicitly.

We evaluate the first two conditional moments of $K$.
\begin{lemma}\label{L2}
Let $\tilde{\mathbf{A}}^{-1}=(b_{ll^\prime})_{l,l^\prime =1, \ldots, p}$. We have
\begin{enumerate}
\item[(1)]
\begin{eqnarray}
M_p&=&\mathit{E}(K|\Omega)\nonumber \\
&=& (\alpha_1 -\alpha_2) \textrm{tr} (\tilde{\mathbf{A}}^{-1}) + \alpha_1 \bar{\mathbf{x}}^{\ast \prime } \tilde{\mathbf{A}}^{-1}\bar{\mathbf{x}}^\ast \nonumber \\
&&- \alpha_2 (\bar{\mathbf{y}}^\ast+ \tilde{\vec{\mu}})^\prime \tilde{\mathbf{A}}^{-1} (\bar{\mathbf{y}}^\ast+ \tilde{\vec{\mu}});\label{e13}
\end{eqnarray}
\item[(2)]
\begin{eqnarray}
B_p^2&=& Var(K|\Omega) \nonumber \\
&=& (\alpha_1 -\alpha_2)^2 (\gamma_x -3) \sum_l b_{ll}^2 + 2(\alpha_1 -\alpha_2)^2 tr(\tilde{\mathbf{A}}^{-2}) + 4\alpha_1^2 \bar{\mathbf{x}}^{\ast\prime} \tilde{\mathbf{A}}^{-2}\bar{\mathbf{x}}^\ast \nonumber \\
&& + 4\alpha_2^2 (\bar{\mathbf{y}}^\ast+ \tilde{\vec{\mu}})^\prime \tilde{\mathbf{A}}^{-2} (\bar{\mathbf{y}}^\ast+ \tilde{\vec{\mu}}) + (4\alpha_1\alpha_2 -4\alpha_2^2)\theta_x \sum_l b_{ll}(\tilde{\mathbf{A}}^{-1} (\bar{\mathbf{y}}^\ast+ \tilde{\vec{\mu}}))_l \nonumber\\
&& - 8\alpha_1\alpha_2 \sum_{ll^\prime} \bar{x}^\ast_l b_{ll^\prime}(\tilde{\mathbf{A}}^{-2} (\bar{\mathbf{y}}^\ast+ \tilde{\vec{\mu}}))_l + (4\alpha_1\alpha_2 -4\alpha_1^2) \theta_x \sum_l b_{ll}(\tilde{\mathbf{A}}^{-1} \bar{\mathbf{x}}^\ast)_l.\label{e14}\nonumber \\
&& \
\end{eqnarray}
\end{enumerate}
\end{lemma}

\paragraph{Proof of Lemma~\ref{L2}.} It is easy to obtain the conditional expectation (\ref{e13}). For the conditional variance of $K$, we first calculate the conditional second moment
\begin{eqnarray*}
\mathit{E}(K^2|\Omega) &=& \mathit{E}_\Omega \Big\{ \alpha_1^2[\mathbf{z}^{\ast \prime} \tilde{\mathbf{A}}^{-1}\mathbf{z}^{\ast} - 2 \bar{\mathbf{x}}^{\ast \prime} \tilde{\mathbf{A}}^{-1}\mathbf{z}^{\ast} + \bar{\mathbf{x}}^{\ast \prime} \tilde{\mathbf{A}}^{-1} \bar{\mathbf{x}}^{\ast}]^2 \\
&& \quad \quad + \alpha_2^2 [\mathbf{z}^{\ast \prime} \tilde{\mathbf{A}}^{-1}\mathbf{z}^{\ast} -2 (\bar{\mathbf{y}}^\ast + \tilde{\vec{\mu}})^\prime \tilde{\mathbf{A}}^{-1}\mathbf{z}^{\ast} +(\bar{\mathbf{y}}^\ast + \tilde{\vec{\mu}})^\prime \tilde{\mathbf{A}}^{-1} (\bar{\mathbf{y}}^\ast + \tilde{\vec{\mu}})]^2 \\
&& \quad \quad-2\alpha_1\alpha_2 [\mathbf{z}^{\ast \prime} \tilde{\mathbf{A}}^{-1}\mathbf{z}^{\ast} - 2 \bar{\mathbf{x}}^{\ast \prime} \tilde{\mathbf{A}}^{-1}\mathbf{z}^{\ast} + \bar{\mathbf{x}}^{\ast \prime} \tilde{\mathbf{A}}^{-1} \bar{\mathbf{x}}^{\ast}] \\
&& \quad \quad \quad \quad \times [\mathbf{z}^{\ast \prime} \tilde{\mathbf{A}}^{-1}\mathbf{z}^{\ast} -2 (\bar{\mathbf{y}}^\ast + \tilde{\vec{\mu}})^\prime \tilde{\mathbf{A}}^{-1}\mathbf{z}^{\ast} +(\bar{\mathbf{y}}^\ast + \tilde{\vec{\mu}})^\prime \tilde{\mathbf{A}}^{-1} (\bar{\mathbf{y}}^\ast + \tilde{\vec{\mu}})] \Big\}.
\end{eqnarray*}
Since
\begin{eqnarray*}
&&\mathit{E}_\Omega\left[\mathbf{z}^{\ast \prime} \tilde{\mathbf{A}}^{-1}\mathbf{z}^{\ast}\right]^2 = (\gamma_x -3) \sum_l b_{ll}^2 + \big(\textrm{tr} \tilde{\mathbf{A}}^{-1}\big)^2 + 2 \textrm{tr} (\tilde{\mathbf{A}}^{-2});\\
&& \mathit{E}_\Omega\left[\mathbf{z}^{\ast \prime} \tilde{\mathbf{A}}^{-1}\mathbf{z}^{\ast}\cdot \bar{\mathbf{x}}^{\ast \prime} \tilde{\mathbf{A}}^{-1}\mathbf{z}^{\ast}\right] = \theta_x \sum_l b_{ll} \big(\tilde{\mathbf{A}}^{-1}\bar{\mathbf{x}}^{\ast}\big)_l;\\
&& \mathit{E}_\Omega\left[ \mathbf{z}^{\ast \prime} \tilde{\mathbf{A}}^{-1}\mathbf{z}^{\ast}\cdot (\bar{\mathbf{y}}^\ast + \tilde{\vec{\mu}})^\prime \tilde{\mathbf{A}}^{-1}\mathbf{z}^{\ast} \right] = \theta_x \sum_l b_{ll} \big(\tilde{\mathbf{A}}^{-1}(\bar{\mathbf{y}}^\ast + \tilde{\vec{\mu}})\big)_l; \\
&& \mathit{E}_\Omega \left[ \bar{\mathbf{x}}^{\ast \prime} \tilde{\mathbf{A}}^{-1}\mathbf{z}^{\ast} \cdot \mathbf{z}^{\ast \prime}\tilde{\mathbf{A}}^{-1} \bar{\mathbf{x}}^{\ast}\right] = \bar{\mathbf{x}}^{\ast \prime} \tilde{\mathbf{A}}^{-2}\mathbf{x}^{\ast};\\
&& \mathit{E}_\Omega \left[  (\bar{\mathbf{y}}^\ast + \tilde{\vec{\mu}})^\prime \tilde{\mathbf{A}}^{-1}\mathbf{z}^{\ast} \cdot \mathbf{z}^{\ast \prime} \tilde{\mathbf{A}}^{-1} (\bar{\mathbf{y}}^\ast + \tilde{\vec{\mu}})\right] = (\bar{\mathbf{y}}^\ast + \tilde{\vec{\mu}})^\prime \tilde{\mathbf{A}}^{-2} (\bar{\mathbf{y}}^\ast + \tilde{\vec{\mu}}),
\end{eqnarray*}
we obtain
\begin{eqnarray*}
\mathit{E}(K^2|\Omega)&=& (\alpha_1 -\alpha_2)^2 (\gamma_x -3) \sum_l b_{ll}^2 + (\alpha_1 -\alpha_2)^2 \big(\textrm{tr} (\tilde{\mathbf{A}}^{-1})\big)^2 + 2 (\alpha_1 -\alpha_2)^2 tr(\tilde{\mathbf{A}}^{-2}) \\
 && + 4\alpha_1^2 \bar{\mathbf{x}}^{\ast\prime} \tilde{\mathbf{A}}^{-2}\bar{\mathbf{x}}^\ast+ 4\alpha_2^2 (\bar{\mathbf{y}}^\ast+\tilde{\vec{\mu}})^\prime \tilde{\mathbf{A}}^{-2} (\bar{\mathbf{y}}^\ast+\tilde{\vec{\mu}}) -8\alpha_1\alpha_2 \bar{\mathbf{x}}^{\ast\prime} \tilde{\mathbf{A}}^{-2} (\bar{\mathbf{y}}^\ast+\tilde{\vec{\mu}})\\
 &&  + 2\alpha_1(\alpha_1-\alpha_2) tr(\tilde{\mathbf{A}}^{-1}) (\bar{\mathbf{x}}^{\ast\prime} \tilde{\mathbf{A}}^{-1}\bar{\mathbf{x}}^\ast) + 2\alpha_2(\alpha_2 -\alpha_1) tr(\tilde{\mathbf{A}}^{-1}) (\bar{\mathbf{y}}^\ast+\tilde{\vec{\mu}})^\prime \tilde{\mathbf{A}}^{-1} (\bar{\mathbf{y}}^\ast+\tilde{\vec{\mu}}) \\
 &&+ 4\alpha_1(\alpha_2 -\alpha_1) \theta_x \sum_l b_{ll} \big(\tilde{\mathbf{A}}^{-1}\bar{\mathbf{x}}^{\ast}\big)_l + 4\alpha_2(\alpha_1 -\alpha_2) \theta_x \sum_l b_{ll} \big(\tilde{\mathbf{A}}^{-1}(\bar{\mathbf{y}}^\ast + \tilde{\vec{\mu}})\big)_l \\
 &&+ \big(\alpha_1 \bar{\mathbf{x}}^{\ast\prime } \tilde{\mathbf{A}}^{-1}\bar{\mathbf{x}}^\ast - \alpha_2 (\bar{\mathbf{y}}^\ast+\tilde{\vec{\mu}})^\prime \tilde{\mathbf{A}}^{-1} (\bar{\mathbf{y}}^\ast+\tilde{\vec{\mu}})\big)^2.
\end{eqnarray*}
Finally, by
\begin{eqnarray*}
Var(K|\Omega)= \mathit{E}(K^2|\Omega) - \mathit{E}^2(K|\Omega),
\end{eqnarray*}
equation (\ref{e14}) follows. The Lemma~\ref{L2} is proved.

The first step of the proof of Theorem~\ref{T1} is similar to the one of the proof of Theorem~\ref{T2} where we ensure that $K-\mathit{E}(K)$ satisfies the Lyapounov condition. The details are referred to (\ref{e14}). Therefore, conditioned on $\Omega$, as $n\to \infty$, the misclassification probability for the D-criterion satisfies
\begin{eqnarray*}
\lim \left\{P_\Omega (2|1) - \Phi \left(\frac{M_p}{B_p}\right)\right\}\to 0.
\end{eqnarray*}
Next, we look for main terms in $M_p$ and $B^2_p$, respectively, using Lemma~\ref{L2}. For $M_p$, we find the following equivalents for the three terms
\begin{enumerate}
\item
\begin{eqnarray*}
(\alpha_1-\alpha_2) \textrm{tr}(\tilde{\mathbf{A}}^{-1}) &=& \frac{p}{n} (\alpha_1 -\alpha_2)\cdot \frac{1}{p} tr(\tilde{\mathbf{S}}^{-1}) \\
&=& \frac{a_1}{n}\cdot \left\{p\left(\frac{1}{n_2+1} -\frac{1}{n_1+1}\right)\right\} + o(\frac{1}{n});
\end{eqnarray*}
\item
\begin{eqnarray*}
\alpha_1 \bar{\mathbf{x}}^{\ast \prime} \tilde{\mathbf{A}}^{-1}\bar{\mathbf{x}}^{\ast} &=& \frac{\alpha_1}{n} \big|\big|\bar{\mathbf{x}}^\ast\big|\big|^2 \cdot \left(\frac{\bar{\mathbf{x}}^\ast}{\big|\big| \bar{\mathbf{x}}^\ast\big|\big|} \right)^\prime \tilde{\mathbf{S}}^{-1} \left(\frac{\bar{\mathbf{x}}^\ast}{\big|\big| \bar{\mathbf{x}}^\ast\big|\big|} \right) \\
&=& \frac{a_1}{n}\cdot \alpha_1 \big|\big|\bar{\mathbf{x}}^\ast\big|\big|^2+o(\frac{1}{n});
\end{eqnarray*}
\item
\begin{eqnarray*}
\alpha_2 (\bar{\mathbf{y}}^\ast + \tilde{\vec{\mu}})^\prime \tilde{\mathbf{A}}^{-1}(\bar{\mathbf{y}}^\ast + \tilde{\vec{\mu}})= \frac{a_1}{n}\cdot \alpha_2 \big|\big|\bar{\mathbf{y}}^\ast + \tilde{\vec{\mu}}\big|\big|^2 + o(\frac{1}{n}).
\end{eqnarray*}
\end{enumerate}
Finally,
\begin{eqnarray}
M_p = \frac{a_1}{n}\cdot \left\{p\left(\frac{1}{n_2+1} -\frac{1}{n_1+1}\right) +\alpha_1 \big|\big|\bar{\mathbf{x}}^\ast\big|\big|^2+ \alpha_2 \big|\big|\bar{\mathbf{y}}^\ast + \tilde{\vec{\mu}}\big|\big|^2\right\} +o(\frac{1}{n}).\label{e15}
\end{eqnarray}

As for $B_p^2$, we find the following equivalents for the seven terms
\begin{enumerate}
\item
\begin{eqnarray*}
&&\left|(\alpha_1-\alpha_2)^2 (\gamma_x -3)\sum_l b_{ll}^2\right| \\
&\leq& \frac{1}{n^2}\left(\frac{1}{n_2+1} -\frac{1}{n_1+1}\right)^2 \big|\gamma_x -3\big| \cdot \textrm{tr}(\tilde{\mathbf{S}}^{-2})\\
 &=& \frac{ya_2}{n^3} \big|\gamma_x -3\big| + o(\frac{1}{n^3}) = O(\frac{1}{n^3});
\end{eqnarray*}
\item
\begin{eqnarray*}
&&2(\alpha_1-\alpha_2)^2 \textrm{tr}(\tilde{\mathbf{A}}^{-2})\\
&=&\frac{2}{n^2}\left(\frac{1}{n_2+1} -\frac{1}{n_1+1}\right)^2 \cdot \textrm{tr}(\tilde{\mathbf{S}}^{-2}) \\
&=& \frac{2ya_2}{n^3} +o(\frac{1}{n^3})= O(\frac{1}{n^3});
\end{eqnarray*}
\item
\begin{eqnarray*}
4\alpha_1^2 \bar{\mathbf{x}}^{\ast \prime} \tilde{\mathbf{A}}^{-2}\bar{\mathbf{x}}^{\ast} = 4\alpha_1^2\frac{a_2 ||\bar{\mathbf{x}}^{\ast}||^2}{n^2} +o(\frac{1}{n^2});
\end{eqnarray*}
\item
\begin{eqnarray*}
4\alpha_2^2 (\bar{\mathbf{y}}^\ast +\tilde{\vec{\mu}})^\prime \tilde{\mathbf{A}}^{-2} (\bar{\mathbf{y}}^\ast +\tilde{\vec{\mu}}) = 4\alpha_2^2 \frac{a_2\big|\big|\bar{\mathbf{y}}^\ast +\tilde{\vec{\mu}}\big|\big|^2}{n_2}+o(\frac{1}{n^2});
\end{eqnarray*}
\item
\begin{eqnarray*}
&&4\alpha_2\big|\alpha_1-\alpha_2\big| \theta_x \sum_l b_{ll} (\tilde{\mathbf{A}}^{-1}(\bar{\mathbf{y}}^\ast +\tilde{\vec{\mu}}))_l \\ &=&\frac{4\alpha_2}{n^2}\left|\frac{1}{n_2+1} -\frac{1}{n_1+1}\right| \sum_l c_{ll} (\tilde{\mathbf{S}}^{-1}(\bar{\mathbf{y}}^\ast +\tilde{\vec{\mu}}))_l \\
&\leq& \frac{4\alpha_2}{n^2}\left|\frac{1}{n_2+1} -\frac{1}{n_1+1}\right|\left(\sum_l c_{ll}^2\right)^{\frac{1}{2}}\cdot \left(\sum_l \left(\tilde{\mathbf{S}}^{-1}(\bar{\mathbf{y}}^\ast +\tilde{\vec{\mu}})\right)_l^2\right)^{\frac{1}{2}}\\
 &\leq& \frac{4\alpha_2}{n^3} \sqrt{p}\cdot \big|\big|\bar{\mathbf{y}}^\ast +\tilde{\vec{\mu}}\big|\big|\sqrt{a_2} + o(\frac{1}{n^2\sqrt{n}});
\end{eqnarray*}
\item
\begin{eqnarray*}
8\alpha_1\alpha_2\sum_{ll^\prime} \bar{x}_l^\ast b_{ll^\prime} (\tilde{\mathbf{A}}^{-2} (\bar{\mathbf{y}}^\ast +\tilde{\vec{\mu}}))_l \leq \frac{8\alpha_1\alpha_2}{n^3}\sqrt{p}\cdot \big|\big|\bar{\mathbf{y}}^\ast +\tilde{\vec{\mu}}\big|\big|\sqrt{a_2}+o(\frac{1}{n^2\sqrt{n}});
\end{eqnarray*}
\item
\begin{eqnarray*}
(4\alpha_1\alpha_2-\alpha_1^2)\theta_x\sum_l b_{ll}(\tilde{\mathbf{A}}^{-1}\bar{\mathbf{x}}^\ast)_l \leq \frac{4\alpha_1}{n^3}\sqrt{p}\cdot ||\bar{\mathbf{x}}^\ast||\sqrt{a_2}+o(\frac{1}{n^2\sqrt{n}}).
\end{eqnarray*}
\end{enumerate}
It can be proved that almost surely,
\begin{eqnarray*}
&& ||\bar{\mathbf{x}}^{\ast}||^2 -\frac{p}{n_1} \to 0,\\
&& \big|\big|\bar{\mathbf{y}}^\ast +\tilde{\vec{\mu}}\big|\big|^2 -\left(\frac{p}{n_2} +\Delta^2\right) \to 0,\\
&& \big|\big|\bar{\mathbf{y}}^\ast +\tilde{\vec{\mu}}\big|\big| - \sqrt{\frac{p}{n_2} +\Delta^2} \to 0.
\end{eqnarray*}
Then the terms 2 and 3 are of order $O(\frac{1}{n^2})$ and 5-7 are of order $o(\frac{1}{n^2})$. Finally,
\begin{eqnarray}
B_p^2 = 4\alpha_1^2\frac{a_2 ||\bar{\mathbf{x}}^{\ast}||^2}{n^2} +4\alpha_2^2 \frac{a_2\big|\big|\bar{\mathbf{y}}^\ast +\tilde{\vec{\mu}}\big|\big|^2}{n^2}+o(\frac{1}{n^2}).\label{e16}
\end{eqnarray}

Since $n_1/n \to \lambda$, we have
 \begin{eqnarray*}
 n_1 \to n\lambda, & n_2 \to n(1-\lambda).
 \end{eqnarray*}
Finally, it holds almost surely,
\begin{eqnarray*}
\lim \left\{\Phi\left(\frac{M_p}{B_p}\right) - \Phi\left(-\frac{\Delta^2}{\sqrt{\frac{y}{\lambda(1-\lambda)}+\Delta^2}} \sqrt{1-y}\right)\right\} \to 0.
\end{eqnarray*}
This ends the proof of Theorem~\ref{T1}.

\subsection{Proof of Theorem~\ref{T2}}
\label{sec:14}
By the assumption 2 in Theorem~\ref{T2}, the covariance matrix is $\mathbf{\Sigma}=\textrm{diag}(\sigma_{ll})_{1\leq l \leq p}$. Under the data-generation models (a) and (b), the misclassification probability (\ref{e10}) can be rewritten as
\begin{eqnarray}
P(2|1)
&=&P\big\{\alpha_1(\mathbf{z}^\ast-\bar{\mathbf{x}}^\ast)^\prime \mathbf{\Sigma} (\mathbf{z}^\ast-\bar{\mathbf{x}}^\ast) -\alpha_2(\mathbf{z}^\ast-\bar{\mathbf{y}}^\ast-\tilde{\vec{\mu}})^\prime \mathbf{\Sigma} (\mathbf{z}^\ast-\bar{\mathbf{y}}^\ast-\tilde{\vec{\mu}}) >0 \big| \mathbf{z}\in \Pi_1\big\}\nonumber\\
&=& P\left(\sum_{l=1}^p k_l>0 \Big|\mathbf{z}\in \Pi_1\right),\label{e17}
\end{eqnarray}
where
\begin{eqnarray*}
k_l=\alpha_1(z^\ast_l-\bar{x}^\ast_l)^2 \sigma_{ll}-\alpha_2(z^\ast_l-\bar{y}^\ast_l -\tilde{\mu}_l)^2 \sigma_{ll}.
\end{eqnarray*}

We firstly evaluate the first two moments of $\sum_{l=1}^p k_l$.
\begin{lemma}\label{L3}
Under the data-generation models (a) and (b), we have
\begin{enumerate}
\item[(1)]
 \begin{eqnarray*}
\mathit{E}(k_l)=-\alpha_2 \sigma_{ll} \tilde{\mu}_l^2,
\end{eqnarray*}
and
\begin{eqnarray}
M_p=\sum_{l=1}^p \mathit{E}(k_l)=-\alpha_2||\vec{\delta}||^2; \label{e18}
\end{eqnarray}

\item[(2)]\begin{eqnarray*}
Var(k_l)=\sigma_{ll}^2\left\{\beta_0 +\beta_1(\gamma) + \beta_2(\theta)\tilde{\mu}_l + 4\alpha_2\tilde{\mu}_l^2\right\},
\end{eqnarray*}
and
\begin{eqnarray}
B_p^2=\sum_{l=1}^p Var(k_l) = \left[\beta_0 +\beta_1(\gamma)\right]\textrm{tr}(\mathbf{\Sigma}^2) +\beta_2(\theta) \mathbf{I}^\prime \Gamma^3\vec{\delta} +4\alpha_2\vec{\delta}^\prime \mathbf{\Sigma}\vec{\delta},\label{e19}
\end{eqnarray}
where
\begin{eqnarray*}
&&\beta_0= \alpha_1^2\frac{6n_1^2+3n_1-3}{n_1^3} +\alpha_2^2\frac{6n_2^2+3n_2-3}{n_2^3}+2(\alpha_1\alpha_2-1),\\
&&\beta_1(\gamma)=\gamma_x\left( \frac{\alpha_1^2}{n_1^3}+(\alpha_1-\alpha_2)^2 \right) +\frac{\alpha_2^2}{n_2^3}\gamma_y,\\
&&\beta_2(\theta) = 4\alpha_2(\alpha_1 -\alpha_2)\theta_x +\frac{4}{n_2^2}\theta_y.
\end{eqnarray*}
If removing the small terms with order $O(p/n_\ast^2)$, then the formula of $B_p^2$ in Theorem~\ref{T2} is obtained.
\end{enumerate}
\end{lemma}

\paragraph{Proof of Lemma~\ref{L3}.} Since $\mathbf{z}^\ast, (\mathbf{x}^\ast_l)$ and $(\mathbf{y}^\ast_l)$ are independent, the variables $(k_l)_{l=1,\ldots,p}$ are also independent. For the expectation of $k_l$, we have
\begin{eqnarray*}
\mathit{E}(k_l) &=& \alpha_1 \sigma_{ll} \cdot \mathit{E}(z^\ast_l-\bar{x}^\ast_l)^2 -\alpha_2 \sigma_{ll} \cdot \mathit{E}(z^\ast_l-\bar{y}^\ast_l-\tilde{\mu}_l)^2\nonumber\\
&=& \alpha_1 \sigma_{ll}\cdot \alpha_1^{-1}- \alpha_2 \sigma_{ll}\cdot (\alpha_2^{-1}+\tilde{\mu}_l^2)\nonumber\\
&=& -\alpha_2\sigma_{ll}\tilde{\mu}_l^2.
\end{eqnarray*}
Equation (\ref{e18}) follows.

For the variance, we have
\begin{eqnarray*}
Var(k_l)&=& \mathit{E}[k_l-\mathit{E}(k_l)]^2 \nonumber \\
&=& \sigma_{ll}^2 \cdot \mathit{E}\left\{\alpha_1 (z^\ast_l-\bar{x}^\ast_l)^2-\alpha_2(z^\ast_l -\bar{y}^\ast_l-\tilde{\mu}_l)^2 +\alpha_2\tilde{\mu}_l^2\right\}^2\nonumber \\[1mm]
&=& \sigma_{ll}^2\cdot \big\{\alpha_1^2 \mathit{E}(z^\ast_l-\bar{x}^\ast_l)^4 +\alpha_2^2 \mathit{E}(z^\ast_l-\bar{y}^\ast_l)^4 + 4 \alpha_2^2\tilde{\mu}_l^2 \mathit{E}(z^\ast_l-\bar{y}^\ast_l)^2 \nonumber \\[1mm]
&&\hskip1cm -2\alpha_1\alpha_2 \mathit{E}\left[(z^\ast_l-\bar{x}^\ast_l)^2 (z^\ast_l-\bar{y}^\ast_l)^2\right] -4 \alpha_2^2 \tilde{\mu}_l \mathit{E}(z^\ast_l -\bar{y}^\ast_l)^3 \nonumber \\[1mm]
&&\hskip1cm + 4\alpha_1 \alpha_2 \tilde{\mu}_l \mathit{E}[(z^\ast_l-\bar{x}^\ast_l)^2(z^\ast_l-\bar{y}^\ast_l)]\big\}.
\end{eqnarray*}
Moreover,
\begin{eqnarray*}
\mathit{E}[z^\ast_l-\bar{x}^\ast_l]^4
&=& \gamma_x\left(1+\frac{1}{n_1^3}\right) +\frac{6n_1^2+3n_1-3}{n_1^3},\\
\mathit{E}[z^\ast_l-\bar{y}^\ast_l]^4 &=& \gamma_x+\frac{\gamma_y}{n_2^3}+\frac{6n_2^2+3n_2-3}{n_2^3},\\
\mathit{E}[z^\ast_l-\bar{y}^\ast_l]^2 &=& \alpha_2^{-1},\\
\mathit{E}[z^\ast_l -\bar{y}^\ast_l]^3 &=& \theta_x -\frac{\theta_y}{n_2^2},\\
\mathit{E}\left\{[z^\ast_l-\bar{x}^\ast_l]^2[z^\ast_l-\bar{y}^\ast_l]^2\right\}
&=& \gamma_x +\frac{1}{\alpha_1\alpha_2}-1,
\end{eqnarray*}
and
\begin{eqnarray*}
\mathit{E}\left\{(z^\ast_l-\bar{x}^\ast_l)^2(z^\ast_l-\bar{y}^\ast_l)\right\}  =\theta_x.
\end{eqnarray*}
Finally, we obtain
\begin{eqnarray*}
Var(k_l)&=& \sigma_{ll}^2 \Bigg\{ \alpha_1^2\left[\gamma_x\left(1+\frac{1}{n_1^3}\right) +\frac{6n_1^2+3n_1-3}{n_1^3}\right] +\alpha_2^2\left[\gamma_x+\frac{\gamma_y}{n_2^3} +\frac{6n_2^2+3n_2-3}{n_2^3}\right]\\
&& \quad \quad + 4\alpha_2^2\tilde{\mu}_l^2\alpha_2^{-1} -2\alpha_1\alpha_2 \left[\gamma_x +\frac{1}{\alpha_1\alpha_2}-1\right] + 4\alpha_1\alpha_2\tilde{\mu}_l \theta_x -4\alpha_2^2 \tilde{\mu}_l \left[\theta_x-\frac{\theta_y}{n_2^2}\right] \Bigg\}\\
&=& \sigma_{ll}^2 \Bigg\{ \gamma_x\left(\alpha_1^2+\frac{\alpha_1^2}{n_1^3}+\alpha_2^2 - 2\alpha_1\alpha_2\right) +\frac{\alpha_2^2\gamma_y}{n_2^3} + \alpha_1^2\frac{6n_1^2+3n_1-3}{n_1^3}\\
&& \quad \quad +\alpha_2^2\frac{6n_2^2+3n_2-3}{n_2^3}-2
+ 4\alpha_2\tilde{\mu}_l^2 +2\alpha_1\alpha_2
 + 4\alpha_2(\alpha_1-\alpha_2)\tilde{\mu}_l \theta_x + \frac{4\tilde{\mu}_l}{n_2^2}\theta_y \Bigg\}\\
&=& \sigma_{ll}^2\Big\{\beta_0 +\beta_1(\gamma) + \beta_2(\theta)\tilde{\mu}_l + 4\alpha_2\tilde{\mu}_l^2\Big\}.
\end{eqnarray*}
Equation (\ref{e19}) follows. Then $B_p^2$ can be rewritten as
\begin{eqnarray*}
B_p^2&=&\Big[ \frac{6n_1+3}{(n_1+1)^2} +\frac{6n_2+3}{(n_2+1)^2} -\frac{2}{n_1+1} -\frac{2}{n_2+1} + \frac{2}{(n_1+1)(n_2+1)} -\frac{3}{n_1(n_1+1)^2} -\frac{3}{n_2(n_2+1)^2}\\
&& \quad +\frac{\gamma_x}{(n_1+1)^2} +\frac{\gamma_y}{(n_2+1)^2} -\frac{2\gamma_x}{(n_1+1)(n_2+1)} +\frac{\gamma_x}{n_1(n_1+1)^2} +\frac{\gamma_x}{n_2(n_2+1)^2}\Big] \textrm{tr}(\mathbf{\Sigma}^2)\\
&&+\left[ 4\frac{n_2}{n_2+1}\left(\frac{1}{n_2+1}-\frac{1}{n_1+1}\right)\theta_x +\frac{4}{n_2^2}\theta_y \right]\mathbf{1}_p^\prime \Gamma^3 \vec{\delta}\\
&& + 4\frac{n_2}{n_2+1}\vec{\delta}^\prime \mathbf{\Sigma}\vec{\delta}\\
&\approx& \Big[ \frac{4}{n_1}+\frac{4}{n_2}+\frac{3}{n_1^2} +\frac{3}{n_2^2} +\frac{2}{n_1n_2} -\frac{3}{n_1^3} -\frac{3}{n_2^3} + \frac{\gamma_x}{n_1^2} +\frac{\gamma_y}{n_2^2} -\frac{2\gamma_x}{n_1n_2} +\frac{\gamma_x}{n_1^3} -\frac{\gamma_y}{n_2^3}\Big]\textrm{tr}(\mathbf{\Sigma}^2)\\
&& + \left[4\left(\frac{1}{n_2}-\frac{1}{n_1}\right)\theta_x +\frac{4}{n_2^2}\theta_y\right]\mathbf{1}_p^\prime \Gamma^3 \vec{\delta}\\
&& + 4\left(1-\frac{1}{n_2}\right)\vec{\delta}^\prime \mathbf{\Sigma}\vec{\delta}.
\end{eqnarray*}
Only keep the terms with order $O(p)$ and $O(p/n_\ast)$ we can get the formula of $B_p^2$ in Theorem~\ref{T2}. The Lemma~\ref{L3} is proved.

We know that $\left[k_l-\mathit{E}(k_l)\right]_{1\leq l \leq p}$ are independent variables with zero mean. We use the Lyapounov criterion to establish a CLT for $\sum_l \left[k_l-\mathit{E}(k_l)\right]$, that is, there is a constant $b>0$ such that
\begin{eqnarray*}
\lim_{p\to \infty} B_p^{-(2+b)}\sum_{l=1}^p \mathit{E}\left[ \big|k_l-\mathit{E}(k_l)\big|^{2+b} \right] \to 0.
\end{eqnarray*}
Since
\begin{eqnarray*}
\big|k_l-\mathit{E}(k_l)\big| &=& \sigma_{ll}\big| \alpha_1 (z^\ast_l -\bar{x}^\ast_l)^2 -\alpha_2(z^\ast_l-\bar{y}^\ast_l)^2 + 2\alpha_2 \tilde{\mu}_l (z^\ast_l-\bar{y}^\ast_l) \big|\\
&\leq& \sigma_{ll}\left\{ \big|z_l^\ast -\bar{x}_l^\ast\big|^2 + \big|z_l^\ast-\bar{y}_l^\ast\big|^2 +2\big|\tilde{\mu}_l\big| \big|z_l^\ast -\bar{y}_l^\ast\big| \right\}\\
&\leq& \sigma_{ll}\left\{  \big|z_l^\ast -\bar{x}_l^\ast\big|^2 +2\big|z_l^\ast-\bar{y}_l^\ast\big|^2 +  \big|\tilde{\mu}_l\big|^2 \right\}\\
&\leq& \sigma_{ll}\left\{ 2\left( \big|z_l^\ast\big|^2 +\big|\bar{x}_l^\ast\big|^2 \right) +4 \left( \big|z_l^\ast\big|^2 +\big|\bar{y}_l^\ast\big|^2  \right) + \big|\tilde{\mu}_l\big|^2 \right\}\\
&\leq& \sigma_{ll}\left\{ 6\left( \big|z_l^\ast\big|^2 +\big|\bar{x}_l^\ast\big|^2 + \big|\bar{y}_l^\ast\big|^2 \right) + \big|\tilde{\mu}_l\big|^2 \right\},
\end{eqnarray*}
the $(2+b)-$norm of $\left[k_l-\mathit{E}(k_l)\right]$ is
\begin{eqnarray*}
||k_l-\mathit{E}(k_l)||_{2+b} &\leq& \sigma_{ll} \left\{ 6\left[ \Big|\Big| |z_l^\ast|^2 \Big|\Big|_{2+b} + \Big|\Big| |\bar{x}_l^\ast|^2\Big|\Big|_{2+b} + \Big|\Big| |\bar{y}_l^\ast|^2 \Big|\Big|_{2+b} \right] + \big|\tilde{\mu}_l\big|^2\right\}\\
&=&  \sigma_{ll} \left\{ 6\left[ \left( \mathit{E}\big|z_l^\ast\big|^{4+b^\prime} \right)^{\frac{1}{4+b^\prime}} +\left( \mathit{E}\big|\bar{x}_l^\ast\big|^{4+b^\prime} \right)^{\frac{1}{4+b^\prime}} +\left( \mathit{E}\big|\bar{y}_l^\ast\big|^{4+b^\prime}\right)^{\frac{1}{4+b^\prime}} \right] + \big|\tilde{\mu}_l\big|^2 \right\}\\
&\leq& \sigma_{ll} \left\{ 6\left[2\gamma_{4+b^\prime, x}^{1/(4+b^\prime)} +\gamma_{4+b^\prime, y}^{1/(4+b^\prime)}\right] + \big|\tilde{\mu}_l\big|^2 \right\}.
\end{eqnarray*}
Then
\begin{eqnarray*}
\mathit{E}\left[k_l-\mathit{E}(k_l)\right]^{2+b} \leq c_b \sigma_{ll}^{2+b} \cdot \left\{ 1 +\big|\tilde{\mu}_l\big|^{4+b^\prime} \right\},
\end{eqnarray*}
where $c_d$ is some constant depending on $b$.
Therefore, as $B_p^2 \approx 4\vec{\delta}^\prime \mathbf{\Sigma}\vec{\delta}=4 \sum_{l=1}^p \tilde{\mu}_l^2 \sigma_{ll}^2$,
\begin{eqnarray*}
B_p^{-(2+b)} \sum_{l=1}^p \mathit{E}[k_l-\mathit{E}(k_l)]^{2+b} &\leq& c_b \cdot \frac{\sum_l \sigma_{ll}^{2+b} +\sum_l \sigma_{ll}^{2+b}|\tilde{\mu}_l|^{4+2b}}{\left( \sum_l \sigma_{ll}\delta_l^2 \right)^{1+b/2}}\\
&=& c_b\cdot \frac{\sum_l \sigma_{ll}^{2+b} +\sum_l \delta_l^{4+2b}}{(\sum_l \sigma_{ll}\delta_l^2)^{1+b/2}} \quad \to 0,
\end{eqnarray*}
by the assumption 4 in Theorem~\ref{T2}. Finally, we have
\begin{eqnarray*}
B_p^{-1} \sum_{l=1}^p\left[k_l-\mathit{E}(k_l)\right]\Rightarrow N(0,1), \ as \  p\to \infty, \ n_\ast \to \infty.
\end{eqnarray*}
This ends of the proof of Theorem~\ref{T2}.




\end{document}